\documentclass[10pt]{article}
\usepackage{graphicx}
\usepackage{dcolumn}
\usepackage{bm}
\usepackage{amsmath}
\usepackage{amsfonts}
\usepackage{natbib}
\begin{document}

\begin{titlepage}
\begin{center}
{\bf Modelling the Bid and Ask Prices of Illiquid CDSs} \\
	\vspace{12pt}
 {Michael B. Walker}\footnote{Department of Physics, University of Toronto, Toronto, ON
M5S 1A7, CANADA; email: walker@physics.utoronto.ca.\label{coords}}$^,$\footnote{I would like to thank Tomasz Bielecki, Jeremy Staum, Dan Rosen, and other participants of the Fields Institute Thematic Program on Quantitative Finance, Toronto, January-June, 2010, as well as John Chinneck, Ken Jackson, Roy Kwon, and Olvi Mangasarian for helpful discussions. Special thanks is also given to an anonymous referee for significant comments encouraging greater detail and further exploration of some issues, and for suggestions on presentation. The support of the Natural Sciences and Engineering Research Council of Canada is acknowledged.\label{ack}} $^,$\footnote{Electronic version of an article published as [IJTAF Vol. 15, Iss. 06, 2012, Page 1250045 (37 pages)] [DOI No: 10.1142/S0219024912500458 ] [copyright World Scientific Publishing Company] [http://www.worldscinet.com/ijtaf/]}\\
\vspace{12pt}
Received: 22 April 2010 	  \\ 
Accepted: 30 January 2012
\end{center}
\vspace{1 cm}


\begin{abstract}
CDS (credit default swap) contracts that were initiated some time ago frequently have spreads and/or maturities that are not available on the current market of CDSs, and are thus illiquid. This article introduces an incomplete-market approach to valuing illiquid CDSs that, in contrast to the risk-neutral approach of current market practice, allows a dealer who buys an illiquid CDS from an investor to determine ask and bid prices (which differ) in such a way as to guarantee a minimum positive expected rate of return on the deal.  An alternative procedure, which replaces the expected rate of return by an analogue of the Sharpe ratio, is also discussed.  The approach to pricing just described belongs to the good-deal category of approaches, since the dealer decides what it would take to make an appropriate expected rate of return, and sets the bid and ask prices accordingly.  A number of different hedges are discussed and compared within the general framework developed in the article.  The approach is implemented numerically, and example plots of important quantities are given.  The paper also develops a useful result in linear programming theory in the case that the cost vector is random.

\vspace{1cm} \noindent keywords:  credit defaults swaps, CDSs,
hedging, valuation, incomplete markets
\end{abstract}
\end{titlepage}

\section{Introduction}\label{introduction}
Illiquid CDS contracts have termination dates and/or spread payments that differ from the termination dates and spread payments of the contracts on the current liquid CDS market.  These illiquid contracts can only be approximately hedged in terms of portfolios of liquid CDS contracts.  This means that the CDS market is incomplete. The purpose of this article is to describe a new approach which takes into account the incompleteness of the CDS market in the valuation procedure, and which gives bid and ask prices that are, in general, different.

A credit default swap (CDS) is a credit derivative that provides insurance against the loss of notional of a corporate bond on default.  The details of CDS contracts are described in section~\ref{market}, as are the details of the CDS market.\footnote{Some readers may find it helpful to read section~\ref{market} before proceeding.}

The current market practice for valuing illiquid CDSs is to use an approach that first determines a risk-neutral measure by calibration to prices of CDSs on the liquid market. A price for an illiquid CDS is then estimated as the expected value of the future payoffs of the illiquid CDS under this risk-neutral measure. In the special case that there is a CDS on the current market having the same time to maturity $T_M$ as that of the illiquid CDS, the value of the illiquid CDS obtained by this procedure is given by
\begin{equation}
	u_M^{Old} =  u_M + (w_M-w_M^{Old})\times RPV01(T_M)
\label{stdCDSValue}
\end{equation}
where $u_M$ in the upfront payment that is made to acquire a long-protection CDS contract with a time to maturity of $T_M$ years (measured from today) on the current market at at spread of $w_M$; $w_M^{Old}$ is the spread of the illiquid contract.  This approach, including a detailed discussion of the calculation of $RPV01(T_M)$ in (\ref{stdCDSValue}), is described in \citet{beu09}, on page 12 of \citet{bnp04}, in \citet{fel06}, and in Chapters 6 and 7 of \citet{oka08}. Short-comings of this approach are that the bid and ask prices are equal, both being given by $u^{Old}$, and that no account is taken of the risks associated with the fact that the illiquid CDS can not be perfectly hedged.

The discussion of the approach of this article begins in section~\ref{NAB} which describes how to hedge an illiquid CDS with CDSs from the market using the hedge that enforces the no-arbitrage bounds on the bid or ask price of the illiquid CDS, and describes the statistical properties of the hedged position.  This is followed, in section~\ref{goodDealBounds}, with a description of how to find good-deal bounds on the bid and ask prices for the illiquid CDS. Sections \ref{NAB} and \ref{goodDealBounds}, which form the heart of the paper, have their own more detailed introductions.

An appendix gives a straightforward extension of the \cite{man79} theorem on uniqueness in linear programming to the case where the cost vector is random, showing that in this case, if there is a solution, then the solution is unique.  A consequence is that the possibility of non-unique solutions need not be considered in the problems studied in this article (or in other case where the cost vector is random).

The literature on the application of incomplete market ideas to problems in the valuation of illiquid derivatives is too large to review in this article focused on CDSs. Fortunately, a detailed and thorough recent review, from which the author has profited, is available in \citet{sta08}.  An elementary introduction to incomplete market ideas, sufficient for an understanding of the present article, can be found in \citet{pli97}. Articles on good-deal bounds for incomplete-market problems include \citet{ber00,car01,cer02,coc00,sta04}. \citet{bie04a,bie04b} describe approaches to pricing defaultable claims in incomplete markets.  However, there is nothing in the literature which describes how to use incomplete-market ideas to obtain bid and ask prices (with a non-zero bid-ask spread) for illiquid CDSs.  In summary, the general ideas which form the basis for this article are well-known from the theory of incomplete markets, so that the originality of the presentation comes from finding a simple way of implementing these ideas for the case of CDSs. An inovation that might be of interest beyond the CDS problem tackled here is the use of a risk measure that tends to zero risk as the arbitrage-free bounds are approached.

The effects of a relatively small degree of illiquidity in determining the bid-ask spreads in the liquid market is not discussed here. \citet{bri10} give a recent survey of this literature.

\section{The CDS Market and Illiquid CDSs}\label{market}
In April 2009, in what is known as the CDS Big Bang, certain changes to the standard CDS contract were introduced.   Pre-CDS-Big-Bang contracts were priced according to a running par spread with zero upfront payment.  Post-CDS-Big-Bang contracts are priced according to an upfront payment with a standardized running spread.  In North America, there is liquidity in contracts with spreads having the two standard values of $w = 100$ bp/yr (for investment-grade names), and $w = 500$ bp/yr (for high-yield names). In Europe, the standardized spread values are $w = 25, 100, 500, 1000$ bp/yr.  For a detailed discussion of post-CDS-Big-Bang contracts see \citep{mar09,mar09b}. 

Standard ISDA CDS contracts have termination dates of March 20, June 20, September 20, and December 20 for any given year (e.g. see \citet{mar09}).  The CDS spread payments are made quarterly on these dates.  Typically, for all contracts concluded during the quarter ending on March 20th of a given year, there is liquidity only in contracts which have a termination date of March 20th some integral number of years in the future, and there is no liquidity in contracts terminating on one of the other standardized termination dates, except, perhaps, for a contract termination date of September 20, 6 months in the future. On March 21, there is a quarterly roll at which time the liquid contracts become those with a termination date of June 20.  Furthermore, the number of annual maturities with liquidity is limited.  For example, for a reasonbly liquid  name, liquidity might be available in contracts with a time to maturity of 1,2,3,5,7 and 10 years. 

It is now clear that, for a given reference name at a given time during a particular quarterly roll, there is liquidity in contracts of a limited number of termination dates, and of a single standardized spread.  Other contracts concluded at previous times (called seasoned CDS contracts) may have termination dates and/or spreads that are not equal to those of the currently liquid contracts just described.  The spreads of these other contracts could be different from those currently on the market for two reasons: a) the contracts could be legacy contracts concluded in the pre-CDS-Big-Bang era at a par spread determined by the market, and b) the contract could be post-CDS-Big-Bang North American contract concluded some time ago at a standardized spread of 100 bp/yr when the name was investment grade; if this name is today considered to be in the high yield category, liquidity might exist only in the 500 bp/yr spreads.
Thus, seasoned CDS contracts could well be illiquid, which will occur when there is no contract on the current market having the same spread and termination date as the seasoned contract. Thus, it will be assumed in this article that investors and dealers can hold two essentially different CDS contracts, those that are currently on the liquid market, and those that are illiquid.  This is admittedly an idealization, since even contracts that are considered to be on the market can be more or less liquid, depending on the name and the maturity, and dealers might have an inventory of illiquid CDSs.

The holder of a CDS contract for a given name that is on the currently liquid CDS market can easily and accurately hedge this contract simply by purchasing the offsetting contract, which will also be on the current market.  On the other hand, the holder of an illiquid CDS contract will generally be unable to purchase an offsetting contract.  Thus, the hedging of an illiquid contract will be carried out by purchasing a portfolio containing (at least one) liquid CDS contracts, as well as a bank deposit.  Furthermore, the hedging of an illiquid contract will be only approximate, so that the hedger's hedged postion is risky, i.e. it has a  realized present value that is uncertain (depending on the default time and recovery rate, both of which are random variables).
This difference in the hedging procedures for liquid and illiquid CDS contracts must be reflected also in different valuation procedures.  Liquid CDSs have values determined by the market. On the other hand, there is no market price for an illiquid CDS.  As noted in Section~\ref{introduction}, this article develops a framework for establishing good-deal bounds for the bid and ask prices of an illiquid CDS.   

\section{No-Arbitrage Hedging Portfolios and their Statistical Properties}\label{NAB}

This section develops a procedure for the static hedging of an illiquid CDS using multiple CDSs from the market together with a cash deposit.  This procedure determines both the no-arbitrage bounds on the bid and ask prices of the illiquid CDS, and the hedging portfolios that enforce these bounds.  These hedging portfolios are the ones that will be used in this article to hedge illiquid CDSs, and their determination and statistical properties are described in depth in this section, which contains numerous technical details and numerical examples.

Subsection~\ref{notation} introduces a number of definitions and notations that will be used in developing the model (which treats both the default time $\tau$ and the recovery rate $\rho$ as continuous random variables).  

The optimization procedure used to determine the hedging portfolios is described in subsection~\ref{staticHedge}.  It is here that the quantity $\Delta(\tau,\rho)$, the realized present value of the payoff stream of the hedged illiquid CDS position when the default time is $\tau$ and the recovery rate is $\rho$, is introduced. The constraint that $\Delta(\tau,\rho)$ be non-negative for all $\tau$ and $\rho$ plays an important role in the determination of the no-arbitrage bounds on the bid and ask prices of the illiquid CDS, as well as the hedging portfolio.  Also, once the hedging portfolio is known, the statistical properties of $\Delta(\tau,\rho)$ (with $\tau$ and $\rho$ as random variables) can be found, and these  determine the risk of the hedged position, and  play an essential role in determining the good-deal bounds on the bid and ask prices.

A particular discretization of the continous variables $\tau$ and $\rho$ is described in subsection~\ref{discretization}, where it is used to convert the optimization problem determining the hedging portfolio to a standard linear programming problem (easily soluble using commercially available software).  The physical probability measure used to compute the statistical properties of the hedged illiquid CDS positions is described in subsection~\ref{physicalMeasure}.  Subsection~\ref{scaling} describes some useful scaling relationships (for the probability density for $\Delta$, for example) that hold when the market has a CDS of the same maturity as that of the illiquid CDS.  Numerical examples illustrating computational details include the determination of no-arbitrage bounds and hedging portfolios (subsection~\ref{NABEx}), and the calculation of the probability densities for $\Delta$ for a single CDS (subsection~\ref{singleCDS}), for the plain vanilla hedge (subsection~\ref{plainVanilla}), and for the multi-CDS hedge (subsection~\ref{DeltaGammaDelta}).

\subsection{Notational Definitions}\label{notation}
The goal will be to construct useful hedging portfolios for an illiquid CDS contract from CDSs of maturities currently on the market, together with an initial cash deposit called $\beta$.  The time at which the hedged position is constructed by purchasing CDSs on the market is denoted by t =0, and the times of the quarterly spread payments made subsequent to $t=0$ are denoted by $T_i,\ i = 1,2,\dots $.  Also, define $T_{i=0}=0$.  The illiquid CDS is charaterized by giving its maturity $T_M$, its spread $w^{Old}$, and its notional $\alpha^{Old}$, which will often be taken to be $+1$ (long protection) or $-1$ (short protection).  The CDSs selected from the market for a given hedging portfolio are labelled by the index $p = 1,2,...,K$, in order of increasing maturity. The maturity of the $p$-th CDS is $T_{n(p)}$, and $n(p)$ is the number of premium payment times to maturity for this CDS. The upfront market price of unit notional of a long position in the $p$-th CDS is $u_p$ and the spread payment is $w_p$.  Allowing both the upfront payment $u_p$  and the spread payment $w_p$ to depend on the CDS maturity $T_{n(p)}$ gives a formulation of the problem that is applicable to both  pre-CDS-Big-Bang and post-CDS-Big-Bang contracts.  For the pre-CDS-Big-Bang contracts, $u_p = 0$, whereas for the post-CDS-Big-Bang contracts $w_p = w$, independent of $p$.   The notional of the $p$-th CDS in the hedge  is called $\alpha_p$; a long (i.e. long protection) CDS position is described by $\alpha_p >0$, whereas $\alpha_p<0$ describes a short CDS position.  The total CDS notional present in the hedging portfolio at its inception is $\alpha_{total} = \sum_{p=1}^K \alpha_p$. 

Let $N = max\{n(K),M\}$. Then $T_N$ is the maximum maturity of all CDSs in the hedged position, which includes all market CDSs in the hedging portfolio, as well as the illiquid CDS.  

Consider the $p$-th long CDS contract from the current market, which has spread $w_p$ and maturity $T_{n(p)}$.  The present value of the spread payment made by the contract holder at time $T_i$, provided no default has occurred up to and including time $T_i$, will be called $g_{p,i}$.  Also, if default occurs at time $\tau$ such that $\tau \in (T_{i-1},T_i]$\, the present value of the loss payment made to the contract buyer minus the spread payment made by the contract buyer at the default time $\tau$  is called $h_{p,i}(\tau,\rho)$. These two present values are given explicitly, per unit notional, as 
\begin{equation}
	g_{p,i} = w_p(T_i-T_{i-1})d_i\delta_{i \le n(p)}, \ \ h_{p,i}(\tau,\rho) = (1-\rho -
	    w_p(\tau - T_{i-1}))d(\tau)\delta_{i\le n(p)}.
	\label{ghDef}
\end{equation}
Here, $d_i \equiv \exp(-r_FT_i)$ and $d(\tau) = \exp(-r_F\tau)$ are discount factors, and $\delta_{i \le n(p)}$ is unity if $i \le n(p)$ and zero otherwise.  The quantities $g_i^{Old}$ and $h_i^{Old}$ relating to the illiquid CDS are similarly defined except that $w^{Old}$ replaces $w_p$, and $M$ replaces $n(p)$.

\subsection{Static Hedging of an Illiquid CDS}\label{staticHedge}
Today's market value of a hedging portfolio constructed, as described above, from K CDSs on the market having notionals $\alpha_p$, $p = 1,...,K$,  together with the initial bank deposit $\beta$, is 
\begin{eqnarray}
	V &=& c^T v, \nonumber \\
	c^T &=& [u_1\ u_2\ \dots\ u_K\ 1], \nonumber  \\
	v^T &=& [\alpha_1\ \alpha_2\ \dots\ \alpha_K\ \beta].
	\label{V}
\end{eqnarray}
where a superscript T, as in $c^T$, indicates the transpose of an array.  The array $c$ is called the cost vector and the array $v$ defines the hedging portfolio.

Now consider the various possible payoff streams for the $p$-th long-protection market-CDS contract of unit notional, maturity $T_{n(p)}$ and spread $w_p$.  The payoff stream $(\tau,\rho)$ is completely characterized by giving its default time $\tau$ and \ recovery rate $\rho$. The present value of the payoff stream ($\tau, \rho)$ for the $p$-th market CDS is
\begin{eqnarray}
	\Delta_p(\tau,\rho)& = & h_{p,I(\tau)}(\tau,\rho) -
		\sum_{k=1}^{I(\tau)-1} g_{p,k},\ \ 
		\tau \in (0,T_{n(p)}],\nonumber \\
	\Delta_p(\tau,\rho) & = & -\sum_{i = 1}^{n(p)} g_{p,k},\ \ 
	\tau > T_{n(p)}.
	\label{DeltaSingle}
\end{eqnarray}
where the function $I(\tau) = i$ when $\tau \in (T_{i-1},T_i)$,  and the quantities $h$ and $g$ are given in (\ref{ghDef}).  The convention $\sum_{k = 1}^0 = 0$ is used in the first of these equations. The present value of the payoff stream $(\tau,\rho)$ for a hedged illiquid CDS of notional $\alpha^{Old}$ is
\begin{equation}
	\Delta(\tau,\rho) = \beta + \alpha^{Old}\Delta^{Old}(\tau,\rho)
		+ \sum_{p=1}^K \alpha_p \Delta_p(\tau,\rho).
	\label{DeltaPort}
\end{equation}

The minimum cost of a hedging portfolio that gives a hedged position with non-negative present values for all payoff streams is determined by the procedure
\begin{eqnarray}
	&&V_{min} = \min(c^T v) \nonumber \\
	\text{subject to}:\ &&\Delta(\tau,\rho) \geq 0,\ \text{for all}\ \tau     > 0 \ \text{and}\ \rho \in [0,1].
\label{hedgePortCont}
\end{eqnarray}
If a dealer takes over an illiquid CDS contract of notional $\alpha^{Old}$ from an investor, the dealer can charge the investor the amount $V_{min}$ and hedge the illiquid CDS so that there is no possibility of a loss.  The dealer will thus make an arbitrage profit. (The case that a perfectly offsetting position to the illiquid CDS can be constructed from market securities is excluded because in that case the so-called illiquid CDS would not be illiquid). The solution to \ref{hedgePortCont}, if it exists, can be found by linear programming procedure of the following subsection, and is unique (see Appendix \ref{uniqueness}).

\subsection [Discretization of Paths and a Discretized Linear Programming Problem]{Discretization of Paths and a Discretized Linear Programming Problem\footnote{This subsection can be skipped on first reading.}}\label{discretization}
In order to carry out the procedure (\ref{hedgePortCont}) using commercially available linear programming software, it is important to replace paths identified by the continuous variables $\tau$ and $\rho$ by a set of discretized paths. This can be done simply and efficiently if the discount factor $d(\tau)$ in the quantity $h_{p,i}(\tau,\rho)$ defined in equation (\ref{ghDef}) is replaced by $d(\tau = 0.5(T_{i-1}+T_i))$.  For $r_F = 2\%$ and $T_i-T_{i-1} = 0.25$, the maximum error in this approximation is  $0.5r_F(T_{i-1}-T_i) = 0.25\%$, and the error averaged over this interval is $(r_F(T_i-T_{i-1}))^2/24 =  2.6\times 10^{-5}\%$.  With this approximation, $\Delta(\tau,\rho)$ is a linear function of $\tau$ and $\rho$ in the rectangle defined by $\tau \in (T_{i-1},T_i]$ and $\rho \in [0,1]$, and can be written
\begin{equation}
	\Delta(\tau,\rho) = n^Tx + C_1,\ \ n^T = [n_1,\ n_2],\ \ 
	x^T = [\tau,\ \rho] 
	\label{DeltaLinear}
\end{equation}
within this rectangle, where $n_1,\ n_2$ and $C_1$ are constants.  Note that if $\Delta(\tau,\rho)$ is non-negative on the four corners of the rectangle, it will be non-negative within the entire rectangle.

Now consider default into states $j = 1,2,\dots,J=4$ of interval $i$, representing the four corners of the above-mentioned rectangle, where the J default states are defined by
\begin{eqnarray}
	j &=& 1 \Rightarrow (\tau,\rho)  = (T_{i-1}+0^+,0),\nonumber \\
	j &=& 2 \Rightarrow (\tau,\rho) =  (T_i,0),\nonumber \\
	j &=& 3 \Rightarrow (\tau,\rho) =  (T_{i-1}+0^+,1),\nonumber \\
	j &=& 4 \Rightarrow (\tau,\rho) =  (T_i,1).
	\label{defaultState}
\end{eqnarray}
Here, $0^+$ is a positive infinitesimal that is taken to zero.
The argument of the previous paragraph has shown that if $\Delta(\tau,\rho) >=0$ for the J default states of a given interval $i$, then it  is true for all $\tau$ and  $\rho$, when $\tau$ is in the interval $i$.  The constraint in the procedure (\ref{hedgePortCont}) will therefore hold if it is imposed for the $J$ default states in each interval $i$, for $i = 1,2,\dots,N$, and for $\tau > T_N$.  Thus, in the following, it will be necessary only to considered the discretized paths labelled by index $i = 1,2,\dots,N$ and index $j = 1,2,\dots J$, together with the path corresponding to $\tau > T_N$. 

The present value of the future payoffs for unit notional of the $p$-th market long-CDS contract that defaults into default state $j$ of interval $i$ can be found from (\ref{DeltaSingle}), and will be called $\Delta_{i,j;p},\ i = 1,2,\dots,N,\ j = 1,2,\dots,J,\ p = 1,2,\dots,K$. Next, the indices $i,j$ will be combined into the single index $k$ by the definition $\Delta_{k,p} = \Delta_{i,j;p}$ where $k = J(i-1)+j$.  Also, to cover the case $\tau > T_N$, define $\Delta_{k = JN+1,p} = \Delta_p(\tau > T_N,\rho=1)$, which is independent of $\tau$ and $\rho$ (in the region $\tau > T_N$).  Finally, the index p is  extended to allow it to have the value $K+1$, so that $\Delta_{k,K+1} = 1, k = 1,2,\dots,NJ+1$ gives the present value of `unit notional' (taken to be 1 dollar) of the initial bank deposit. The column vector $\Delta^{Old}_k, k = 1,2,\dots,NJ+1$ is defined similarly to any of the $\Delta_{k,p}$ having $p \le K$, and gives the present values of the discretized payoff streams for the illiquid CDS.  Finally, to have a clear distinction between the arrays giving the present values of the discretized payoff streams of the market CDSs, and the corresponding array for the illiquid CDS, define the array B such that $B_{k,p} = \Delta_{k,p}$ for all $k$ and $p$, and the column vector $b$ such that $b_k = \Delta_k^{Old}$ for all $k$.
 
Now suppose that an investor holding unit notional of an illiquid short-protection CDS ($\alpha^{Old} = -1$ in (\ref{DeltaPort})) approaches a dealer with a request that the dealer take over this short-protection position.  To assess the profitability of such an acquisition the dealer can solve what will be called the primal linear programming problem given by the left-hand column of
\begin{equation}
\begin{array}{ll}
\text{PRIMAL} & \text{DUAL}\\
V^{(+)} = \min(c^Tv) & V^{\prime(+)} = \max(b^Tx)\\
\text{subject to:}\ Bv \geq b\ \ \ \ \ \ \  &\text{subject to:}\ B^Tx = c \\
v\ \ \text{unrestricted} &	x \geq 0.
\label{primalDualLUB}
\end{array}
\end{equation}
The notation $V^{(\sigma)}$ is used in (\ref{primalDualLUB}) and in (\ref{primalDualGLB}) where $\sigma = +$ (or $\sigma = -$) refers to the hedged short-protection (or long-protection) illiquid CDS.
The corresponding dual linear programming problem is also given here.  The concept of duality and its consequences are topics treated in many books on linear programming.
The functions $c^Tv$ of $v$, and $b^Tx$ of $x$ are called the objective functions of the respective problems.  The conditions $Bv \geq b$, and $B^T x = c$ together with $x \geq 0$, are called the constraints.  Optimization is carried out with respect to the variables $v$ and $x$, respectively.  A solution of a primal problem for $v$ gives a hedging portfolio that enforces the non-arbitrate bounds.  A solution of the dual problem for $x$ gives a linear pricing measure.
The primal problem defined here is a discretized version of the problem stated in procedure (\ref{hedgePortCont}).

An investor who transfers a short-protection illiquid CDS contract to a dealer as just described is effectively buying protection, corresponding to the terms of the illiquid contract, from the dealer.  The price paid by an investor to a dealer when buying protection is called the ask price.  Suppose the dealer charges the investor the amount $V^{(+)}$ for protection and then uses this amount to buy the hedging portfolio.  The dealer will then make an arbitrage profit.  If the dealer charges the investor less than $V^{(+)}$, then the dealer will not have the possibility of making an arbitrage profit using the hedge just described.  The price $V^{(+)}$ is thus the least upper bound (LUB) on the arbitrage-free range of ask prices, and the corresponding hedging portfolio $v^{(+)}$ is called the LUB hedging portfolio.  Also, note the definition $v^{(+)T} = [\alpha_1^{(+)}\ \alpha_2^{(+)}\ \dots\ \alpha_K^{(+)}\ \beta^{(+)}] = [\alpha^{(+)},\beta^{(+)}]$ of the LUB hedging portfolio. 

The above discussion has considered an illiquid short-protection ($\alpha^{Old} < 0$ in (\ref{DeltaPort})) CDS of unit notional ($|\alpha^{Old}| = 1$). For a short-protection CDS of arbitrary notional ($|\alpha^{Old}|$ arbitrary) multiply the quantities $V^{(+)}$, $v^{(+)}$, and $b$ of the primal problem in (\ref{primalDualLUB}) by $|\alpha^{Old}|$.

Now consider the case of an investor who owns an illiquid long-protection CDS contract of unit notional ($\alpha^{Old} = +1$ in (\ref{DeltaPort})), and who asks a dealer to take over the contract.  Note that this involves a change of sign of $\alpha^{Old}$ with respect to the problem considered immediately above. A change of sign  defined by $\alpha = -\tilde{\alpha}$ and $\beta = -\tilde{\beta}$ is also introduced into (\ref{DeltaPort}) so that the hedging portfolio in this case is defined by $\tilde{v}^T =[\tilde{\alpha}^T \tilde{\beta}]$.  The dealer's hedged position described by (\ref{DeltaPort}), is now interpreted as a long-protection illiquid position and a short position in the hedging portfolio $\tilde{v}$. The constraints in the primal linear programming problem defined by the left hand column of 
\begin{equation}
\begin{array}{ll}
\text{PRIMAL} & \text{DUAL}\\
V^{(-)} = \max(c^T\tilde{v}) & V^{\prime(-)} = \min(b^Tx)\\
\text{subject to:}\ B\tilde{v} \leq b\ \ \ \ \ \ \  &\text{subject to:}\ B^Tx = c \\
\tilde{v}\ \ \text{unrestricted} &	x \geq 0
\label{primalDualGLB}
\end{array}
\end{equation}
guarantee that the dealer's hedged position has only non-negative payoffs.  The amount realized by the dealer in shorting the hedging portfolio $\tilde{v}$ is $V^{(-)}$.  If this amount is paid to the investor for the long-protection illiquid contract, the dealer will have constructed a hedged position having only non-negative payoffs at zero net cost.  The amount $V^{(-)}$ is thus the greatest lower bound (GLB) on the arbitrage-free range of bid prices for the illiquid long CDS and $\tilde{v}$ is called the GLB hedging portfolio.  (The bid price is the amount paid by the dealer to the investor for protection.)

The feasible region of a linear programming problem is the set of points ($v$ or $x$ above) that satisfy the constraints.
According to the duality theorem, if either the primal or the dual problem has a feasible solution with a finite objective value, then the other problem has a feasible solution with the same objective value.  This means that if the dual LUB problem has an optimal solution with a finite objective value $V^{\prime(+)}$, then the primal problem does also, with $V^{(+)}=V^{\prime(+)}$.  Similarly for the GLB problem, which gives $V^{(-)}=V^{\prime(-)}$.

Note that the feasible region is the same for both the LUB and GLB dual linear programming problems.  Therefore, if optimal solutions exist for both problems, the optimal objective values will satisfy
\begin{equation}
	V^{(-)} \le V^{(+)}.
\end{equation} 

\subsection{The Physical Measure}\label{physicalMeasure}
The holder of the hedged position wishing to proceed to an estimate of its statistical properties must establish physical measures describing holder's views, as a result of detailed research on the question, of the probability density for the default time $\tau$, (called $\Upsilon(\tau)$) and the probability density for the recovery rate $\rho$ conditional on default at time $\tau$ (called $\gamma(\rho|\tau)$). The conditioning of $\gamma$ on $\tau$ is useful for conducting scenario analyses.  However, in the examples in this article, the effects of this conditioning will not be studied, $\tau$ and $\rho$ will be treated as independent variables, and $\gamma(\rho|\tau) = \gamma(\rho)$.

In the numerical examples discussed in this article, the default probability density is assumed to be specified in terms of a constant hazard rate $h$, so that
\begin{equation}
	\Upsilon(\tau) = h\exp(-h\tau), \ 0 \leq \tau \leq T_N\ \ \text{and} \ \ \Upsilon_0 = \exp(-hT_N)
	\label{Upsilon}
\end{equation}
where $\Upsilon_0$ is the probability of survival to time $T_N$.
Also, the hazard rate is given as
\begin{equation}
	h = -log(1-PD_1);
	\label{hazardRate}
\end{equation}
where $PD_1$ is an estimated probability for default within the first year.	A deterministic time-dependent hazard rate could easily be incorporated into the method of this article, but this will not be discussed in detail.

\begin{figure} [t!] 
\includegraphics[scale = 0.6]{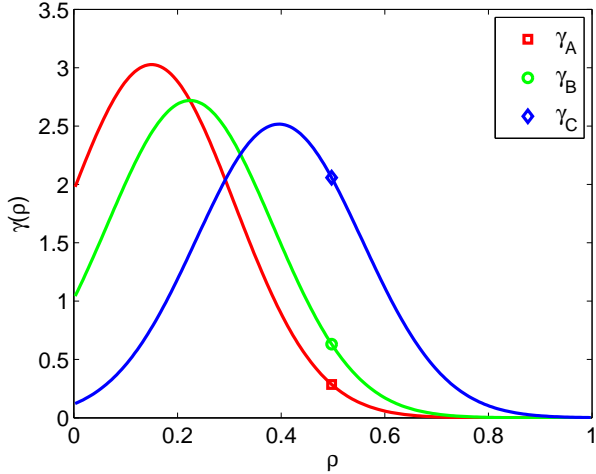} 
\caption{Plot of three different probability densities for the recovery rate $\rho$. The probability densities $\gamma_A(\rho)$, $\gamma_B(\rho)$ and $\gamma_C(\rho)$ are defined on the interval $0 \leq \rho \leq 1$, and on this interval are proportional to  normal probability densities having means of 0.15, 0.224 and 0.396 respectivelly, with all having standard deviations of 0.16 .  The mean values of $\rho$ over the interval $\rho \in [0,1]$ for the three probability densities are $\overline{\rho}_A = 20$ \%,  $\overline{\rho}_B = 25$ \%, and $\overline{\rho}_C = 40$ \%.}
\label{recoveryRateDensityABC}
\end{figure}

Fig.~\ref{recoveryRateDensityABC} shows three different probability densities for the recovery rate, $\gamma_A(\rho)$, $\gamma_B(\rho)$, and $\gamma_C(\rho)$, used in the numerical examples below. The probability density $\gamma_A(\rho)$ was used in most of the examples, while $\gamma_B(\rho)$ and $\gamma_C(\rho)$ are used in section \ref{stability} to test the robustness of the solutions with respect to changes in $\gamma(\rho)$.

Given the probability densities described above, the mean present value of the possible payoff streams for the hedged positions can be evaluated as 
\begin{equation}
	\overline{\Delta} = \int_0^{T_N} d\tau \int_0^1 d\rho
		\Delta(\tau,\rho) \Upsilon(\tau)\gamma(\rho)
	 + \Upsilon_0 \Delta_0.
	 \label{DeltaBar}
\end{equation}
where $\Delta_0$ is the present value of the payoff stream when the default time exceeds the longest CDS maturity in the hedged position.

It is also possible to evaluate the probability density $\Gamma(\Delta)$ of the possible payoff stream present values $\Delta$.

\subsection{Scaling in a Special Case}\label{scaling}
This subsection considers the special case in which there is a market CDS, say the $p_M$-th, for which the maturity is equal to the maturity of the illiquid CDS, i.e. for which $T_{n(p_M)} = T_M$. For this case, it will be shown that certain quantities scale in a definite way with the variable $W = |w_{p_M}-w^{Old}|$.

In (\ref{DeltaPort}) take the notional of the illiquid CDs to be $\alpha^{Old} = -(\sigma 1)$ where $\sigma = +$ ($\sigma = -$) corresponds to unit notional of a short-protection (long-protection) CDS.  Also set $\alpha_{p_M} = (\sigma 1)+\delta\alpha_{p_M}$.  Then $\Delta(\tau,\rho)$ of (\ref{DeltaPort}) can be written as
\begin{eqnarray}
	\Delta(\tau,\rho) &=& \beta 
		+ \sum_{p \neq p_M}\alpha_p \Delta_p(\tau,\rho) +
		\delta\alpha_{p_M}\Delta_{p_M}(\tau,\rho)
		-(\sigma 1)(w_{p_M} - w^{Old}){\cal T}_M(\tau), \nonumber \\
	  {\cal T}_M(\tau) &=&  \sum_{k = 1}^{I(\tau)-1} (T_k-T_{k-1})d_k
	 	+ (\tau - T_{I(\tau)-1})d(\tau),\ \ 
	 	\tau \in (0,T_M], \nonumber \\
	{\cal T}_M(\tau) &=& {\cal T}_{M,0} \equiv \sum_{k = 1}^{n(p_M)} (T_k-T_{k-1})d_k, 
	 	\ \ \tau > T_M.
	\label{DeltaPort2}
\end{eqnarray}	
Note that the letter ${\cal T}$ in the calligraphic type style is different from the italic letter $T$. It can be seen that ${\cal T}_M(\tau)$ is a continuous function of $\tau$ defined for all $\tau \geq 0$. 
The usefulness of this new expression for $\Delta(\tau,\rho)$ is that the spread $w^{Old}$ of the illiquid CDS appears only in the combination $(w_{p_M} - w^{Old})$ explicitly shown, and nowhere else.  Now define
\begin{equation}
	\beta^\prime = \frac{\beta}{W};\ \ \alpha_p^\prime = 
	\frac{\alpha_p}{W}, p \neq p_{M};\ \
	\alpha_{P_M}^\prime = \frac{\delta\alpha_{p_M}}{W};\ \ \Delta^\prime(\tau,\rho) = \frac{\Delta(
	\tau,\rho)}{W}.
	\label{vPrimeDef}
\end{equation}
(The quantity $W$ was defined at the beginning of this subsection.)
The quantity $\Delta^\prime(\tau,\rho)$ is now found to be
\begin{equation}
	\Delta^\prime(\tau,\rho) = \beta' + \sum_{p=1}^K \alpha^\prime_p 
\Delta_p(\tau,\rho) -\sigma \mu{\cal T}_M(\tau),
\label{DeltaPrime}
\end{equation}	
where $\mu = sgn(w_{p_M} - w^{Old})\ (=\pm)$.
The procedure (\ref{hedgePortCont}) can now be written as
\begin{eqnarray}
	&&V_{min}^\prime = \min(c^T v') \nonumber \\
	\text{subject to}:\ &&\Delta'(\tau,\rho) \geq 0,\ \text{for all}\ \tau     > 0 \ \text{and}\ \rho \in [0,1].
\label{hedgePortContReduced}
\end{eqnarray}
where $v^{\prime T} = [\alpha_1^\prime, \alpha_2^\prime,\dots,\alpha_K^\prime,\beta^\prime].$
A solution of (\ref{hedgePortContReduced}) gives numerical values the components of $v^\prime$, now called $v^{\prime (\sigma\mu)}$, which depend on the value of the product $\sigma \mu$, but not on $W$. 

Once a solution of (\ref{hedgePortContReduced}) has been obtained for a given $\sigma \mu$, the no-arbitrage bounds on the ask and bid prices can be found from
\begin{equation}
V^{(\sigma) (\mu)}(W) = (\sigma 1)W(c^T v^{\prime[\sigma\mu]}) + u_{p_M}.
\label{VBidAskScale}
\end{equation}
Note that $V^{(\sigma)(\mu)}(W) - u_{p_M}$ varies linearly with $W$.

The notation $V^{(\sigma)}$ was introduced just following (\ref{primalDualLUB}), and the notation $V^{(\sigma) (\mu)}$ and $v^{\prime[\sigma\mu]}$ was used just above.  Thus, the quantities $V^{(+)}$, $V^{(+)(-)}$ and $v^{\prime[+]}$ can all occur. In $V^{(+)}$, the quantity + appears between parenthuses $()$ and respresents a value of $\sigma$.  In $V^{(+)(-)}$, there are two successive sets of parentheses: the quantity + in the first set represents a value of $\sigma$, while the quantity $-$ in the second represents a value of $\mu$.  In $v^{\prime[+]}$, the quantity + in the brackets [] represents a value of the product $\sigma\mu$.

Let $\Delta^{\prime[\sigma\mu]}(\tau,\rho)$ and $\Delta^{[\sigma\mu]}(\tau,\rho)$ be the values of $\Delta^{\prime}(\tau,\rho)$  and $\Delta(\tau,\rho)$, respectively, when $v^{\prime} = v^{\prime [\sigma\mu]}$.  Furthermore, let $\Delta^{*\prime[\sigma\mu]}$ and $\Delta^{*[\sigma\mu]}$ be the random variables corresponding to $\Delta^{\prime[\sigma\mu]}(\tau,\rho)$ and $\Delta^{[\sigma\mu]}(\tau,\rho)$ when $\tau$ and $\rho$ are taken to be random variables.  Then $\Delta^{*[\sigma\mu]} = W \Delta^{*\prime[\sigma\mu]}$.  Also, for $W$ having a given value, if $\Delta^{*\prime[\sigma\mu]}$ has the value $\Delta^\prime$, then $\Delta^{*[\sigma\mu]}$ has the value $W\Delta^\prime$, and if $\Delta^{*\prime[\sigma\mu]}$ has the value $\Delta^\prime + d\Delta^\prime$ then $\Delta^{*[\sigma\mu]}$ has the value $W(\Delta^\prime + d\Delta^\prime) = \Delta + d\Delta$, where $d\Delta = W d\Delta^\prime$. Now, if $\Gamma^{[\sigma\mu]}(\Delta,W)d\Delta$ is the probability that $\Delta^{*[\sigma\mu]} \in (\Delta,\Delta+d\Delta]$ when $W$ has the constant value indicated by the second argument, and $P^{[\sigma\mu]}(\Delta^\prime) d\Delta^\prime$ is the probability that $\Delta^{*\prime [\sigma\mu]} \in (\Delta^\prime, \Delta^{\prime} + d\Delta^\prime]$, then
\begin{equation}
	 \Gamma^{[\sigma\mu]}(\Delta,W)d\Delta =
	  P^{[\sigma\mu]}(\Delta^\prime) d\Delta^\prime =
	  P^{[\sigma\mu]}\left(\frac{\Delta}{W}\right) \frac{d\Delta}{W}.
	  \label{GammaFromP}
\end{equation}
From this relation, it follows that the probability density $\Gamma^{[\sigma\mu]}(\Delta,W)$ satisfies the scaling relation
\begin{equation}
	\Gamma^{[\sigma\mu]}(\frac{\Delta}{f},\frac{W}{f}) = f\Gamma^{[\sigma\mu]}(\Delta,W).
\label{GammaScale}	
\end{equation}
where $f$ is an arbitrary positive number.  This result determines the probability density $\Gamma^{[\sigma\mu]}(\Delta,W)$ for arbitrary positive values of $W$ in terms of that for a single positive value of $W$. Note, for example, that the changes in the probability density $\Gamma(\Delta)$ when $W$ is reduced by a factor of 2 can be desribed roughly as a compression of $\Gamma$ by a factor of 2 towards the $\Delta = 0$ axis, and an expansion by a factor of 2 away from the $\Gamma = 0$ axis, in plots such Fig.~\ref{figGammaMulti} below.

Finally, it follows from (\ref{vPrimeDef}) that
\begin{equation}
	\overline{\Delta}^{[\sigma\mu]} = W\overline{\Delta}
	^{\prime [\sigma\mu]},
	\label{deltaBar}
\end{equation}
and thus that $\overline{\Delta}^{[\sigma\mu]}$ varies linearly with $W$.

\subsection{No-Arbitrage Bounds and Hedging Portfolios}\label{NABEx}

\begin{table}
\begin{tabular}{|c||c|c|c|c|c|c|}
\hline
    maturity &1&2&3&4&5&7 \\ \hline \hline
    par spreads (bp) &1068&1241&1282&1253&1219&1149\\ \hline \hline
    upfront\ (\%) & 5.25 &12.47 &18.08 &21.56 &24.05 &27.00 \\ \hline  
\end{tabular}
\caption{The given par spreads for General Motors Corporation senior CDSs on 20 March 2008 are as quoted by Thomson Datastream.  These par spreads were converted to upfronts by using the ISDA CDS Standard Model as described in \citet{isd09} and in \citet{beu09}, and taking the  running spread and the recovery rate to have the standard values of $w = 500$ bp/yr and $\rho = 20$\% generally used for high-yield names with this model.  Upfronts obtained in this way were used for the numerical examples of this report since  recent upfront quotes from the market were not available to the author.  General Motors filed for bankrupcy protection on June 1, 2009. The CDS recovery rate, determined by auction, was $\rho =$ 12.5\% \citep{reu09}.} \label{upfronts}
\end{table} 

This section describes numerical results for the no-arbitrage bounds and for the LUB and GLB hedging portfolios for an illiquid CDS for which the reference name is General Motors.  The upfront prices of the market CDSs that are used in constructing the hedging portfolio are given in Table~\ref{upfronts}. Since other numerical results that make use of the same set of numerical input parameters will be given below, these parameters will be stated once and for all in Table~\ref{SPL}, and this list of parameters will be called the standard parameter list. A given numerical example does not necessarily use all of these parameters.  For example, the numerical example of this subsection uses only the parameters specified in the first eight rows of the table, except that $w^{Old}$ is considered as a variable parameter.

\begin{table}
\begin{tabular}{|l|l|}
\hline
Quantity & Value \\ \hline 
maturity of illiquid CDS & $T_M = 5$ years from today \\
notional of illiquid CDS & unity \\
running spread of illiquid CDS & $w^{Old} = 100$ bp/yr \\
maturities of market CDSs & 1, 2, 3, 4, and 5 years \\
upfront prices of market CDSs & as given in Table~\ref{upfronts} \\
running spread of market CDSs & $w_p = w = 500$ bp/yr for all maturities \\
quarterly spread payment interval & $T_i-T_{i-1} = 0.25$ yr, $i = 1,2,\dots$ \\
risk-free interest rate & $r_F = 2 \%$ \\
probability of default within 1 year & $PD_1 = 30\%$ \\
recovery rate probability density & $\gamma_A$
    of Fig.~\ref{recoveryRateDensityABC} \\
expected return on capital at risk & $R_T = 25 \%$  \\  
\hline
\end{tabular}
\caption{The standard parameter list.}
\label{SPL} 
\end{table} 

\begin{figure} [t!] 
\includegraphics[scale = 0.6]{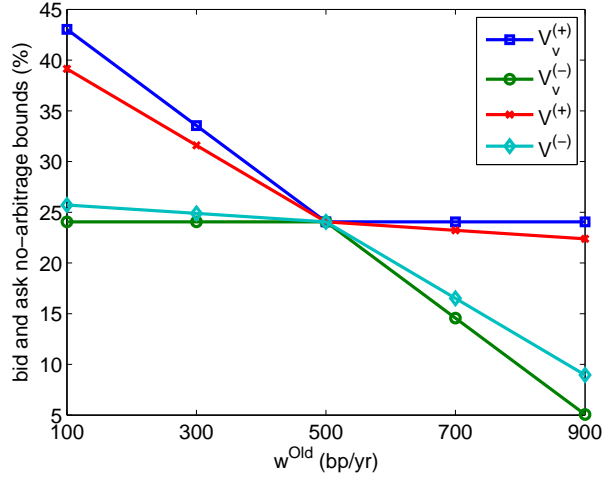}
\caption{Plot of bid and ask no-arbitrage bounds for an illiquid CDS versus $w^{Old}$.  $V^{(\sigma)}$ refers to the multi-CDS hedge, while $V^{(\sigma)}_V$ refers to the vanilla, single-CDS hedge.  Note that the separation between the upper (ask) bound and the lower (bid) bound is greater for the vanilla hedge than it is for the multi-CDS hedge.  The data used to obtain this plot was taken from Table~\ref{SPL}.}
\label{VVsWOld}
\end{figure}
  
Numerical results for two different sets of LUB and GLB hedging portfolios have been obtained by following the procedures (\ref{primalDualLUB}) and (\ref{primalDualGLB}) for the primal problem, and are presented in Fig.~\ref{VVsWOld}.  For the first case considered, the current CDS market will be assumed to consist of a single CDS having the same maturity as that of the illiquid CDS to be hedged, and the GLB and LUB portfolios constructed under this assumption will be called vanilla hedging portfolios.  This is similar to a common industry practice (e.g. see p. 98 of \citet{oka08}) of hedging with a single CDS of the same notional and maturity from the market, except that a bank deposit has been added to turn the hedging portfolio into an LUB or GLB portfolio, and the primal optimization procedures~(\ref{primalDualLUB}) and (\ref{primalDualGLB}) are allowed to determine both the bank deposit and the notional of the market CDS.  The values of the vanilla LUB and GLB portfolios (i.e. the cost of putting these portfolios together by buying their components on the market) are called $V_v^{(\sigma)}$ (with a subscript $v$ for vanilla). These values are also the no-arbitrage bounds on the price of the illiquid CDS in the given market It turns out that in this case, the optimization procedure finds that the single market CDS in the hedging portfolio has the same notional as that of the illiquid CDS to be hedged, thus leading to perfect cancellation of the loss on default payments from the market CDS and the illiquid CDS.  It is in fact possible to use the result of the preceding sentence, together with elementary arguments, to obtain the analytic results for $V_v^{(\sigma)}$ derived below by scaling arguments, and given explicitly by (\ref{VNABPlainVanilla}).

In the second case considered, the current CDS market will be assumed to consist of maturities of 1, 2, 3, 4, and 5 years as indicated in Table~\ref{SPL}. The LUB and GLB and hedging portfolios constructed in this case are called multi-CDS hedges. The values of the multi-CDS hedges are called $V^{(\sigma)}$.  These values represent the costs of constructing the hedges from their components on that market, and are also the no-arbitrage bounds on the price of the illiquid CDS in the given market. (Also, in contrast to the values of the preceding paragraph, the quantities $V^{(\sigma)}$ do not carry a subscript $v$). 

The results for both cases just discussed are shown in Fig.~\ref{VVsWOld}, and the piecewise linear behaviour observed here can be seen to be consistent with the dependence on $W = |w-w^{Old}|$ and $\mu = sgn(w-w^{Old})$ implied by equation~(\ref{VBidAskScale}).  When $w^{Old} = 500$ bp/yr, the values of all hedging portfolios coincide.  For this value of $w^{Old}$, all GLB and LUB portfolios consist of a single CDS from the market having a maturity of 5 years, unit notional, and a running spread of 500 bp/yr, equal to the running spread of the illiquid CDS, and the hedging is perfect. In fact, for this particular value of $w^{Old}$, the supposedly illiquid CDS is on the liquid market. Also, for a given value of $\mu = sgn(w-w^{Old})$, $V^{(\sigma)}$ (as well as $V^{(\sigma)}_v$) varies linearly with $W$. Finally, the slope of $V^{(\sigma)}$ (as well as that of $V^{(\sigma)}_v$) versus $w^{Old}$ depends only on the product $\sigma \mu$.

\begin{table}
\begin{tabular}{|c||c|c|c|c|c|c||c|}
\hline
quantity & $\alpha_1$ & $\alpha_2$ & $\alpha_3$ & $\alpha_4$ & $\alpha_5$ &  $\beta$ &$\alpha_{total}$ \\ \hline \hline
$v^{(+)}$ &	-0.0319  & -0.0342  & -0.0368 & -0.0395  & 1.0000 & 0.1720   & 0.8576  	\\ \hline
$\tilde{v}^{(-)}$ & -0.0403 &  -0.0431 & - 0.0462  &  -0.0495 & 1.1791 & 0.0000   & 1.0000 
  \\ \hline
\end{tabular}
\caption{The table gives the notionals for each maturity of the CDS purchased on the market, as well as the initial cash deposit, in the multi-CDS LUB and GLB hedging portfolios. More precisely, $\alpha_p^{(+)},\ p = 1,\dots,K=5$ and $\beta^{(+)}$ are given for the row headed by $v^{(+)}$, and $\tilde{\alpha}_p^{(-)}$ and $\tilde{\beta}^{(-)}$ are given for the row headed by $\tilde{v}^{(-)}$. The sum of the individual notionals $\alpha_p$ in the LUB or GLB portfolio is $\alpha_{total}$.  The illiquid CDS being hedged has maturity $T_M = 5$ years,  spread $w_M^{Old} = 100$ bp/yr, and notional 1.0000.}
\label{tableWOld100} 
\end{table} 

The notionals of all market CDSs present in the multi-CDS hedging portfolios when the illiquid CDS has a spread of $w^{Old} = 100$ bp/yr (as it would have been if the illiquid CDS contract was a North American post-CDS-Big-Bang contract concluded at a time when the reference name was investment grade) are given in Table~\ref{tableWOld100}.

Fig.~\ref{VVsWOld} shows that, for any value of $w^{Old}$ (except 500 bp/yr), the value $V_v^{(+)} > V^{(+)}$. Recall that the value $V_v^{(+)}$ is obtained by minimizing this value subject to certain constraints.  When the portfolio over which one is minimizing is extended from a single CDS on the market to include four more market CDSs, one expects to get a lower minimum.  A corresponding remark explains the result that $V_v^{(-)} < V^{(-)}$ at any given $w^{Old}$.  When the market consists of CDSs with maturities 1, 2, 3, 4, 5 yrs, the no-arbitrage bounds on the illiquid CDS price are given by $V^{(\sigma)}$, and the prices $V_v^{(\sigma)}$ are outside the no-arbitrage bounds.  One can still use a vanilla hedge to hedge an illiquid CDS if one wishes, but one should be careful to have the sale price of an illiquid CDS lying within the no-arbitrage bounds that are appropriate for the existing CDS liquid market.

\subsection{Example: $\Delta(\tau, \rho)$ and $\Gamma(\Delta)$ for a Single CDS}\label{singleCDS}
Eventually, the present values of the payoff streams $\Delta(\tau,\rho)$ and the probability density $\Gamma(\Delta)$ for these payoff streams will be evaluted for a portfolio of CDSs.  The discussion in this subsection for a single long-protection CDS will help in understanding the properties a portfolio of CDSs.  The single CDS considered here has a maturity of $T_M = 5$ yr, $M = 20$, a notional of unity, and a spread of $w = 500$ bp/yr.

\begin{figure} [t!] 
\includegraphics[scale = 0.4]{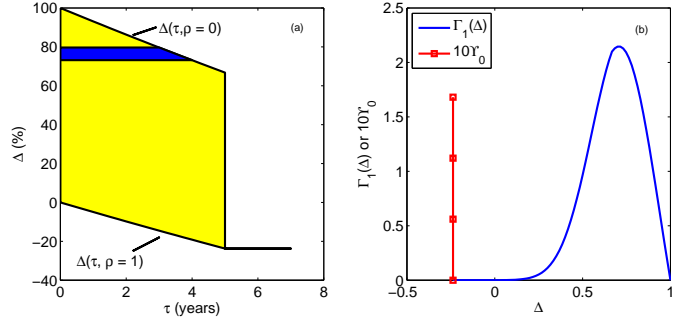}
\caption{The region of the $\Delta-\tau$ plane in which $\Delta(\tau,\rho)$ is defined is shown as the shaded areas (both lightly shaded (yellow) and heavily shaded (blue)) of panel (a) for $\tau \in (0,T_M]$, and as the line given by $\Delta = \Delta_0$ for $\tau > T_M$. The heavily shaded (blue) area of panel (a) is bounded from above by the line $\Delta = \Delta_1 \equiv 80\%$ and from below by $\Delta = \Delta_2 \equiv 73\%.$ (This defines the quantities $\Delta_1$ and $\Delta_2$ used in the evaluation of $\Gamma_1(\Delta)$.) }
\label{figCDSSingle}
\end{figure}

The quantity $\Delta(\tau,\rho = 1)$ represents the spread contribution to $\Delta(\tau,\rho)$, which,  from equations~(\ref{DeltaSingle}) and (\ref{DeltaPort2}) is given explicitly by
\begin{equation}
	\Delta(\tau,\rho = 1) =  -w{\cal T}_M(\tau).
	\label{DeltaSpreadSingleCDS}
\end{equation}
Since ${\cal T}_M(\tau)$ is continuous in $\tau$ (as noted following (\ref{DeltaPort2})), $\Delta(\tau,\rho = 1)$ is also.
For $\tau > T_M$, this gives
\begin{equation}
	\Delta(\tau > T_M,\rho = 1) \equiv \Delta_0 =   -w\sum_{k=1}^M(T_k-T_{k-1})d_k.
	\label{Delta0}
\end{equation}
$\Delta(\tau,\rho = 1)$ is a continuous function of $\tau$ defined for all $\tau > 0$, and, for the parameters assumed in this section, is shown in Fig.~\ref{figCDSSingle}(a).

Also note that the derivative of $\Delta(\tau,\rho = 1)$ when $\tau \in (T_{i-1},T_i]$ is
\begin{equation}
	\frac{d\Delta(\tau,\rho = 1)}{d\tau} = -we^{-r_F\tau}[1-r_F(\tau-T_{i-1})].
	\label{derivative}
\end{equation} 
It follows from this result, with the assumption $T_i-T_{i-1} = 0.25$, that $\Delta(\tau,\rho=1)$ varies monotonically with $\tau$ unless $r_F > 400\%$, i.e. except under the most extreme circumstances.

For $\tau \in (0,T_M]$, $\Delta(\tau,\rho)$ can be formed by adding the loss contribution to the spread contributions, which gives
\begin{equation}
	\Delta(\tau,\rho) = \Delta(\tau,\rho = 1) + (1-\rho)e^{-r_F\tau}
	\label{eqDeltaTauRho}
\end{equation}
From this equation, and from the fact that $\Delta(\tau,\rho = 1)$ is continuous in $\tau$ for $\tau > 0$, it is clear that $\Delta(\tau,\rho)$ is jointly continuous in $\tau$ and $\rho$ in the region $\tau \in (0,T_M],\ \rho \in [0,1]$.  There is, however, except for $\rho = 1$, a discontinuity in $\Delta(\tau,\rho)$ when $\tau$ crosses $T_M$ at fixed $\rho$, since there is no loss contribution when $\tau > T_M$. Note that the loss contribution on the right hand side of (\ref{eqDeltaTauRho}), which is proportional to $(1-\rho)$, is non-negative.  The region of the $\Delta-\tau$ plane in which $\Delta$ is defined is shown as the shaded areas (both lightly shaded (yellow) and heavily shaded (blue)) of Fig.~\ref{figCDSSingle}(a) for $\tau \in (0,T_M]$, and as the line given by equation given by equation~(\ref{Delta0}) for $\tau > T_M$. For a given $\tau \in (0, T_M]$, the loss contribution varies linearly with $\rho$.  Thus, for any given value of $\tau \in (0,T_M]$, the value of $\Delta(\tau,\rho)$ is easily inferred from Fig.~\ref{figCDSSingle}(a) by linear interpolation between $\Delta(\tau,\rho = 1)$ and $\Delta(\tau,\rho = 0)$.     

The final task of this subsection will be to calculate the probability density $\Gamma(\Delta)$, defined above in the paragraph following equation (\ref{DeltaBar}).  The random variable corresponding to the quantity $\Delta(\tau,\rho)$, when $\tau$ and $\rho$ are taken to be random variables, will be called $\Delta^\ast$.  Also, the random variable corresponding to the default time $\tau$ will be called $\tau^\ast$. Let $A$ be the event that  $\Delta^\ast \in (\Delta, \Delta + d\Delta]$, $B_1$ be the event that $\tau^\ast \in (0,T_M]$ and $B_2$ be the event that $\tau^\ast > T_M$.  The events $B_1$ and $B_2$ are mutually exclusive and exhaust all possibilities for $\tau^\ast$.  Therefore, the probability of event $A$ is $P(A) = P(AB_1)+P(AB_2)$ where $P(AB_k)$ is the probability of both $A$ and $B_k$ occurring, $k = 1,2$.  Note that $P(A) = \Gamma(\Delta)d\Delta$ and define $P(AB_k) = \Gamma_k(\Delta)d\Delta$.  This leads to the result
\begin{equation}
	\Gamma(\Delta) = \Gamma_1(\Delta)+\Gamma_2(\Delta).
	\label{GammaSum}
\end{equation}

The quantity $\Gamma_2(\Delta)$ is
\begin{equation}
	\Gamma_2(\Delta) = \Upsilon_0 \delta(\Delta - \Delta_0)
	\label{Gamma2}
\end{equation}
where $\delta(\Delta - \Delta_0)$ is the Dirac delta-function and $\Upsilon_0 = e^{-hT_M}$ is the probability that $\tau^\ast > T_M$.  

To prepare for the calculation of $\Gamma_1(\Delta)$, first note that
\begin{equation}
	P(\Delta^\ast \in (\Delta_1,\Delta_2];\tau^\ast \in (0,T_M]) =
		\int_{S_B} d\tau d\rho \Upsilon(\tau) \gamma(\rho),
		\label{PSB}
\end{equation}
where the integration is over the surface $S_B$ in $(\tau,\rho)$ space such that $\Delta$ and $\tau$ are in the heavily shaded (blue) area of Fig.~\ref{figCDSSingle}(a).   Equation~(\ref{eqDeltaTauRho}) relates the three variables $\Delta,\ \tau,$ and $\rho$. Rather than choosing the variables $(\tau,\rho)$ to be the independent integration variables in (\ref{PSB}) it is convenient to use $(\tau,\Delta)$.  The associated change in the element of surface area is $d\tau d\rho =  d\tau d\Delta/[\partial(\tau,\Delta)/
		\partial(\tau,\rho)] = \exp(r_F \tau)d\tau d\Delta$, where $\exp(r_F \tau)$ is the Jacobian of the transformation. Also make use of the result
\begin{equation}
	\rho(\tau,\Delta) = 1+\Delta(\tau,\rho = 1)e^{r_F\tau}
		-\Delta e^{r_F \tau}
\end{equation}
obtained inverting inverting $\Delta(\tau,\rho)$ of equation~(\ref{eqDeltaTauRho}).  Now shrink the difference $\Delta_2 - \Delta_1$ so that it becomes a positive infinitesimal (effectively, $S_B$ becomes a line at a constant value of $\Delta$) let $d\Delta = \Delta_2 - \Delta_1$, and define $\Delta$ to be the mean of $\Delta_2$ and $\Delta_1$.	This yields the expression
\begin{equation}
	\Gamma_1(\Delta) = \int_{\tau_1(\Delta)}^{\tau_2(\Delta)} \Upsilon(\tau) \gamma[\rho(\tau,\Delta)]e^{r_F \tau} d\tau.
\label{Gamma1Delta}	
\end{equation}
where $\tau_1(\Delta) = 0$ and $\tau_2(\Delta)$ is determined by numerical solution of the equation
$\Delta(\tau_2,\rho = 0) = \Delta$. In a similar manner, $\Gamma_1(\Delta)$ can be obtained for the full range of ${\Delta}'s$ for which it is  nonzero by taking appropriate values for $\tau_1(\Delta)$ and $\tau_2(\Delta)$.  $\Gamma_1(\Delta)$ and $\Upsilon_0$ are shown in panel (b) of Fig.~\ref{figCDSSingle}.

The reason that $\Gamma_1(\Delta)$ goes rapidly to zero as $\Delta$ moves towards more negative values in Fig.~\ref{figCDSSingle}(b) is that $\gamma_A(\rho)$ goes rapidly to zero as $\rho$ moves toward unity (see Fig.~\ref{recoveryRateDensityABC}).  The reason that $\Gamma_1(\Delta)$ goes linearly to zero as $\Delta$ moves towards its maximum value is that the range of $\tau$ over which one integrates in equation~(\ref{Gamma1Delta}) goes linearly to zero as the maximum value of $\Delta$ is approached, while $\gamma(\rho)$ remains nonzero as $\rho$ approaches zero (see Fig.~\ref{figCDSSingle}(a)).

So far in this subsection, it is a single long-protection CDS that has been considered.  The payoffs for the corresponding short-protection CDS are the negatives of those for the long-protection CDS.  Thus, the figure corresponding to Fig.~\ref{figCDSSingle} for a short-protection CDS is obtained from Fig.~\ref{figCDSSingle} by reflecting in the line $\Delta = 0$, for both panel (a) and panel (b).

\subsection{Example: $\Delta(\tau)$ and $\Gamma(\Delta)$ for the Plain Vanilla Hedge}\label{plainVanilla}
Subsection~\ref{NABEx} gave a numerical example of the determination of the hedging portfolio for a vanilla hedge of an illiquid CDS of maturity $T_M = 5$ yr, $M=20$ and spread $w^{Old} = 100$ bp/yr.  The vanilla hedging portfolio by definition consisted of a CDS from the market having the same maturity $T_M$ but a different spread, $w$, together with an initial cash deposit.  The notional of the market CDS and the amount of the initial cash deposit were determined by an optimization procedure. For the numerical example considered, the optimum notional of the CDS from the market was found to be  unit long-protection for the hedging of a unit short-protection illiquid CDS, and unit short-protection for the hedging of the illiquid long-protection CDS.  There may however be sets of input parameters different from those of Table~\ref{SPL} for which the market CDS in the hedging portfolio has a notional different from unity.  

In this subsection, the market CDS also has the same maturity $T_M = 5$ yr, with  $M=20$, as the illiquid CDS to be hedged, and a different spread $w$.  However, the notional of the market CDS is not determined by optimization, but is set equal to unit long-protection for the hedging of a unit short-protection illiquid CDS, and unit short-protection for the hedging of a unit long-protection illiquid CDS. This corresponds to current industry practice and has the attractive feature that the losses on default are offsetting in the hedged position. In addition, a cash deposit is included in the hedging portfolio and is determined in such a way that the hedged portfolios are those that enforce the upper and lower no-arbitrage bounds.  The hedge where the notional of the market CDS is chosen in this way is called the plain vanilla hedge, to distinguish it from the vanilla hedge described earlier, where the notional of the market CDS is determined by optimization. Often, the two procedures will give the same results.

\begin{figure} [t!] 
\includegraphics[scale = 0.4]{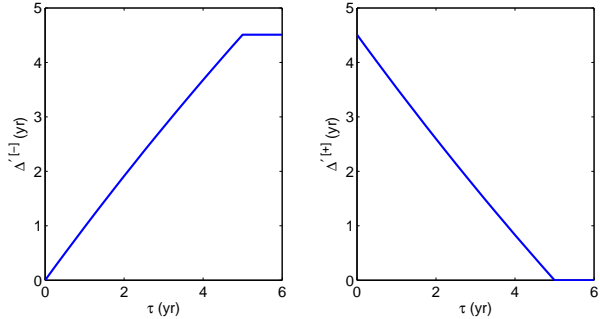}
\caption{The quantities $\Delta^{\prime [\sigma\mu]}$  obtained for the plain vanilla hedged position are plotted vs $\tau$. The maturity of the illiquid CDS is $T_M =$ 5 yr.}
\label{figDeltaVsTauVanilla}
\end{figure}

The problem just described will be solved by using some results from subsection~\ref{scaling} on scaling.  The quantity $\Delta^\prime$  of equation~(\ref{DeltaPrime}) for the present problem is
\begin{equation}
	\Delta^{\prime[\sigma\mu]}(\tau) = \beta^{\prime[\sigma\mu]} -    [\sigma\mu]{\cal T}_M(\tau).
	\label{DeltaPrimePlainVanilla}
\end{equation}
Note that there is no dependence on the recovery rate $\rho$.
The unknown cash deposit $\beta^\prime$ is determined by minimizing $\beta^\prime$ subject to the condition that $\Delta^\prime(\tau)$ is non-negative for all positive default times $\tau$. The results depend on the value of the quantity $\sigma\mu$ and are
\begin{equation}
	\beta^{\prime [+]} = {\cal T}_M(\tau > T_M) \equiv {\cal T}_{M,0},\ \ \beta^{\prime [-]}  = 0.
\label{betaPrime}
\end{equation}
The corresponding values of $\Delta^\prime(\tau)$ are
\begin{equation}
	\Delta^{\prime [+]}(\tau) = {\cal T}_{M,0} - {\cal T}_M(\tau),\ \ 
	\Delta^{\prime [-]}(\tau) = {\cal T}_M(\tau),
	\label{DeltaPrimeResult}
\end{equation}
and the values of $\Delta^\prime_0$ in the case of no default are
\begin{equation}
	\Delta^{\prime [+]}_0 = 0,\ \ \Delta^{\prime [-]}_0 
	= {\cal T}_{M,0}.
	\label{DeltaPrime0}
\end{equation}
The present values of the payoff streams with default time $\tau$ for the hedged illiquid CDSs are 
\begin{equation}
	\Delta^{[\sigma\mu]}(\tau) = W\Delta^{\prime [\sigma\mu]}(\tau).
\end{equation}
The quantities $\Delta^{\prime[\sigma\mu]}(\tau)$ are plotted in Fig.~\ref{figDeltaVsTauVanilla}.  The necessary parameter values are taken from Table~\ref{SPL}.

The no-arbitrage bounds on the bid and ask prices are obtained from (\ref{VBidAskScale}) which, for the plain vanilla hedge, reduces to
\begin{equation}
	V^{(\sigma)(\mu)} = (\sigma 1)W\beta^{\prime [\sigma\mu]} + u
	\label{VNABPlainVanilla}
\end{equation}
where $u$ is the upfront cost of the long-protection market CDS of maturity $T_M$. Explicit results obtained by using (\ref{betaPrime}) in (\ref{VNABPlainVanilla}) are
\begin{equation}
	V^{(+)(+)} = W{\cal T}_{M,0} + u,\ \ V^{(+)(-)} = u,\ \ V^{(-)(+)}   = u,\ \ V^{(-)(-)} = -W{\cal T}_{M,0} + u.
	\label{VNABPlainVanillaExplicit}
\end{equation}
These results agree perfectly with the results shown in Fig.~\ref{VVsWOld}.

The probability densities for $\Delta^{\prime[\sigma\mu]}(\tau)$ will now be calculated.  Consider the case $\sigma\mu = -$. Since all detailed discussion below will consider only this case, write $\Delta^{\prime [-]}(\tau) = \Delta^\prime(\tau) = T_M(\tau)$; i.e., for simplicity, omit the superscript $[-]$ in most of what follows.   As in (\ref{GammaSum})and the paragraph containing it, the probability density $P(\Delta^\prime)$ is written as a sum of two parts
\begin{equation}
	P(\Delta^\prime) = P_1(\Delta^\prime) + P_2(\Delta^\prime)
\end{equation}
where $P_1(\Delta^\prime)$ is the part corresponding to $\tau \in (0,T_M]$, and $P_2(\Delta^\prime)$ is the part corresponding to $\tau > T_M$.  $P_2$ is given by
\begin{equation}
	P_2(\Delta^\prime) = \Upsilon_0 \delta(\Delta^\prime - {\cal T}_{M,0})
	\label{P2}	
\end{equation}
where $\Upsilon_0 = \exp(-h T_M)$. Also,
\begin{equation}
	P_1(\Delta^\prime) =   
		\left. \frac{\Upsilon(\tau)}{|d\Delta^\prime(\tau)/d\tau|}
		\right |_{\tau = \tau(\Delta^\prime)}
		\label{P1PlainVanillaExact}
\end{equation}
Note that $P_1(\Delta^\prime)$ has a non-zero positive value for $\Delta^\prime \in (0,{\cal T}_{M,0}]$ when the derivative $[d\Delta^\prime(\tau)d\tau]_{\tau = \tau(\Delta^\prime)}$ is finite.  At the end points of this interval, $P_1(\Delta^\prime)$ jumps discontinuously to zero. 
A useful approximation is to set $d\Delta^\prime(\tau)/d\tau \approx \exp(-r_F\tau)$ in the interval where it is non-zero. This approximation is equivalent to setting the quantity in the square brackets in (\ref{derivative}) equal to unity.  Thus, (\ref{P1PlainVanillaExact}) simplifies to
\begin{equation}
	P_1^{[-]}(\Delta^\prime) = h\exp(-(h-r_F)\tau(\Delta^\prime)),
	\label{P1Approx}
\end{equation}	
where the fact that $\sigma\mu = -$ has been understood throughout this calculation is now indicated explicitly through the superscript $[-]$. The result (\ref{GammaFromP}) can now be combined with (\ref{P1Approx}) to obtain 
$\Gamma_1^{[-]}(\Delta,W)$.  Similar results can be obtained for $\Gamma_1^{[+]}(\Delta,W)$.

The probability densities $\Gamma_1^{[\sigma\mu]}(\Delta,W)$ versus $\Delta$ for the vanilla hedge are plotted in Fig.~\ref{figGammaMulti} where they are compared with those for the multi-CDS hedge.  The quantity $W = |w-w^{Old}| = 0.04$ for these plots, so $W$ is not shown explicitly as an argument of $\Gamma_1^{[\sigma\mu]}$.

\subsection{Example: $\Delta(\tau,\rho)$ and $\Gamma(\Delta)$ for Multi-CDS Hedge}\label{DeltaGammaDelta}

This subsection gives a numerical example of the calculation of the realized present value function $\Delta(\tau,\rho)$, and of the probability density $\Gamma(\Delta)$, for the illiquid CDS position hedged with a multi-CDS hedge as described in detail in subsection~\ref{NABEx}.  Before doing so, however, a paragraph will be devoted to describing some general properties of the function $\Delta(\tau,\rho)$.

\begin{figure} [t!] 
\includegraphics[scale = 0.4]{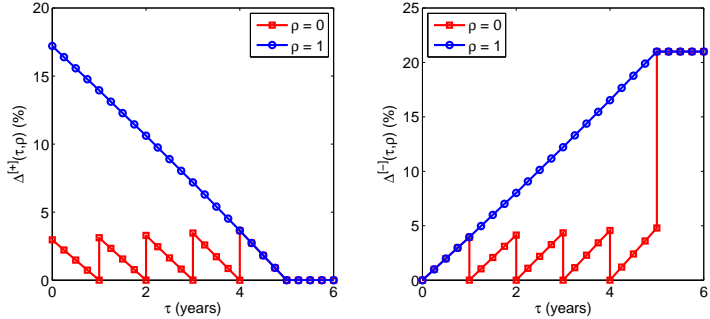}
\caption{Plots of $\Delta^{[\sigma]}(\tau,\rho)$, the realized present value of the payoff stream of the hedged illiquid CDS when default occurs at time $\tau$ and the recovery rate is $\rho$, plotted versus $\tau$.  The lines labelled $\rho =1$ (and $\rho = 0$) are the lines $\Delta^{[\sigma]} = \Delta^{[\sigma]}(\tau,\rho=1)$ (and $\Delta^{[\sigma]} = \Delta^{[\sigma]}(\tau,\rho = 0)$).  The value of $\Delta(\tau,\rho)$ for a given $\tau$ and and arbitrary value of $\rho \in [0,1]$ can be obtained by linear interpolation of $\rho$ between the lines for $\rho = 0$ and $\rho = 1$ for that $\tau$ (e.g. see (\ref{DeltaOfRho})).} 
\label{figDelta} 
\end{figure} 


Because the realized present value function $\Delta^{(\sigma)}(\tau,\rho)$ (the superscript $(\sigma)$ was defined following equation~(\ref{VBidAskScale})) for a hedged illiquid CDS position is a sum of the realized present value contributions from the market CDSs and the illiquid CDS, plus the present value of a cash deposit, its continuity properties can be simply understood from a knowledge of the continuity properties of a single CDS.  These have been described in detail in subsection~\ref{singleCDS}. Because the premium leg $\Delta(\tau,\rho = 1)$ for a single CDS is a continuous function of $\tau$ for all $\tau > 0$, the premium leg $\Delta^{(\sigma)}(\tau,\rho=1)$ for a hedged illiquid CDS postion (which is a sum of contributions from single CDSs) will be also.  Also,  a single CDS of notional $\alpha$ (which is postive for a long-protection CDS), has a value of $\Delta(\tau,\rho) = \Delta(\tau,\rho = 1) + \alpha(1-\rho)$, $\rho \in [0,1]$, for $\tau$ less than or equal to its maturity, and a value of $\Delta(\tau,\rho = 1)$ for $\tau$ greater than its maturity.  Note the discontinuity of $\Delta(\tau,\rho)$ as a function of $\rho$ when $\tau$ is equal to the CDS maturity.  Now, let $\{n;\tau\}$ be a set of indices $n$ labelling the individual CDSs in a given hedged illiquid CDS position that have maturities greater than or equal to the time $\tau$.  Then, 
\begin{equation} \Delta^{(\sigma)}(\tau,\rho) = 
   \Delta^{(\sigma)}(\tau,\rho=1) + (1-\rho)
  \sum_{\{n;\tau\}}  \alpha_n.
  \label{DeltaOfRho}
\end{equation}
Note that $\Delta^{(\sigma)}(\tau,\rho)$ has a discontinuity in the region where it is defined whenever $\tau$ crosses the line $\tau = T_n$, where $T_n$ is the maturity of one of the constituent CDSs (including both the illiquid and market CDSs), except if $\rho = 1$.
Now, consider two neighboring maturities $T_a$ and $T_b > T_a$ occurring in a given hedged illiquid CDS position.  From the preceding discussion, it follows that $\Delta^{(\sigma)}(\tau,\rho)$ 
is defined in the region of the $\Delta^{(\sigma)}-\tau$ plane bounded by the lines $\tau = T_a$, $\tau = T_b$, $\Delta^{(\sigma)} = \Delta^{(\sigma)}(\tau,\rho=1)$ and $\Delta^{(\sigma)} = \Delta^{(\sigma)}(\tau,\rho=0)$.  Furthermore, within this region, $\Delta^{(\sigma)}(\tau,\rho)$ is jointly continuous in $\tau$ and $\rho$ for $\tau \in (T_a,T_b)$ and $\rho \in (0,1)$.  These considerations are also valid when $T_a = 0$ and $T_b$ is the smallest maturity in the hedged position, and when $T_a$ is the largest maturity in the hedged position and $T_b$ is $+\infty$. Finally, note that if $\sum_{\{n;\tau\}}  \alpha_n = 0$ for a given interval $\tau \in (T_a,T_b]$, then $\Delta^{(\sigma)}(\tau,\rho)$ is independent of $\rho$ in that interval and can be written simply as $\Delta^{(\sigma)}(\tau),\ \ \tau \in (T_a,T_b]$.

Now consider the example where the hedged illiquid CDS positions are those determined in subsection~\ref{NABEx}, and for which the relevent data on the notionals of the constituent market CDSs, and on the values of the cash deposits $\beta$, are given in Table~\ref{tableWOld100}. The realized present value functions $\Delta^{[\sigma\mu]}(\tau,\rho)$ for this case are plotted in Fig.~\ref{figDelta}.  These functions can be obtained, for example, from (\ref{DeltaOfRho}).  For a fixed default time $\tau$,  $\Delta^{[\sigma\mu]}(\tau,\rho)$ varies linearly with $\rho$ for $\rho \in [0,1]$. The point respresenting $\Delta^{[\sigma\mu]}(\tau,\rho)$ in Fig.~\ref{figDelta} is thus easily determined by linear interpolation of $\rho$ between the points $\Delta^{[\sigma\mu]}(\tau,\rho=0)$ and $\Delta^{[\sigma\mu]}(\tau,\rho=1)$.   

The maturities of the CDSs that make up the hedged illiquid CDS positions being studied in this subsection are $T_{n(p)},\ \ p = 1,2,\dots,K = 5$, with $T_M = T_{n(K)}$, and $T_{n(p)} = p$ yr, $p = 1,2,\dots,K$.  From the above discussion on continuity, $\Delta(\tau,\rho)$ is continuous inside the region $\tau \in (T_{n(p)-1},T_{n(p)}]$ and $\rho \in [0,1]$, $p = 1,2,\dots,K$. (for simplicity, the superscript $[\sigma\mu]$ is supressed in this paragraph.)  By definition, $T_{n(0)} = 0$. Also, for $\tau > T_M$, $\Delta(\tau,\rho) = \Delta_0 = \text{constant}$.  Now, similarly to (\ref{GammaSum}) and its preceding text, let event $A$ be the event $\Delta^\ast \in (\Delta, \Delta+d\Delta]$, event $B_{1,p}$ be the event $\tau^\ast \in (T_{n(p)-1},T_{n(p)}],\ \ p = 1,2,\dots,K$, and event $B_2$ be the event $\tau^\ast > T_M$.  Thus $P(A) = \sum_{p=1}^K P(AB_{1,p}) + P(AB_2)$.  Now define $P(A) = \Gamma(\Delta)d\Delta$, $P(AB_{1,p}) = \Gamma_{1,p}(\Delta)d\Delta\ \ p = 1,2,\dots,K$ and $P(AB_2) = \Gamma_2(\Delta)d\Delta$. Then
\begin{equation}
	\Gamma(\Delta) = \Gamma_1(\Delta)+\Gamma_2(\Delta),\ \
	\Gamma_1(\Delta) = \sum_{p=1}^K \Gamma_{1,p}(\Delta)
	\label{GammaSumMulti}
\end{equation}	 

Note first of all that
\begin{equation}
	\Gamma^{[\sigma\mu]}_2(\Delta) = \Upsilon_0 \delta(\Delta - 
		\Delta^{[\sigma\mu]}_0)
\end{equation}
where $\Upsilon_0 = e^{-hT_M}$ is the probability of the default time being greater than $T_M$.  (Here $\Upsilon_0 = 16.8\%$, $\Delta^{[+]}_0 = 0.0\%$, and $\Delta^{[-]} = 21.0\%$.)

\begin{figure} [t!] 
\includegraphics[scale = 0.4]{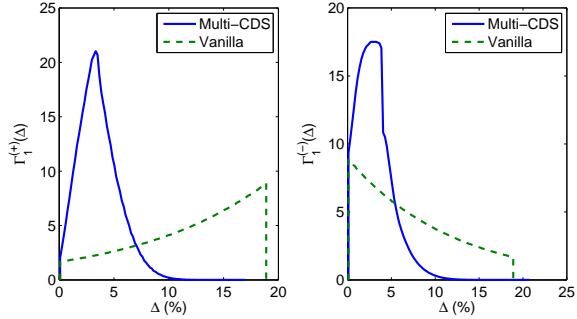}
\caption{Plots of $\Gamma_1^{[\sigma\mu]}(\Delta)$ for both the multi-CDS and vanilla hedges (see legends) for an illiquid CDS of maturity $T_M = 5$ years and spread $w_M^{Old} = 100$ bp/year. The multi-CDS LUB and GLB hedging portfolios are given in Table~\ref{tableWOld100}.  The left panel and the right panel are for the LUB and GLB cases, respectively.  Also, for the LUB case, $\Delta^{[+]}_0 = 0.0\%$ for both the multi-CDS and vanilla hedges, while for the GLB case, $\Delta^{[-]}_0 = 21.0\%$ for the multi-CDS hedge, and 18.9\% for the vanilla hedge.  The probability of $\Delta = \Delta_0$ is $\Upsilon_0 = 16.8 \%$ in all cases.}
\label{figGammaMulti}
\end{figure}

The probability densities $\Gamma^{[\sigma\mu]}_1(\Delta)$ for the multi-CDS case being studied in this subsection are those labelled ``Multi-CDS'' in Fig.~\ref{figGammaMulti}.  The parameter values necessary for this calculation are given in Table~\ref{SPL}.  A qualitative understanding of certain detailed features of this figure follows from the discussion of the computational details given in subsections~\ref{singleCDS} and \ref{plainVanilla}. These features include the fact that $\Gamma^{[\sigma\mu]}_1(\Delta)$ has a non-zero positive value at $\Delta = 0$, the fact that $\Gamma^{[+]}_1(\Delta)$ has a discontinuity when $\Delta = \Delta^{[+]}(\tau = 4\ \text{yr})$ (which is barely visible), the fact that $\Gamma^{[-]}_1(\Delta)$ has a discontinuity when $\Delta = \Delta^{[-]}(\tau = 1\ \text{yr})$ (which is more clearly visible), and the fact that $\Gamma^{[+]}_1(\Delta)$ and $\Gamma^{[-]}_1(\Delta)$ both approach zero rapidly for values of $\Delta$ greater than 5\%.

One can now compare the effectiveness of the multi-CDS hedge with that of the plain vanilla hedge from the point of view of reducing the riskiness of the hedged position. For the hedged short-protection illiquid CDS, both the multi-CDS and the plain vanilla hedges find $\Delta^{[+]}_0 = 0.0\%$ occurs with a probability of 16.8\%.  However, the overall spread in the values of $\Delta$ that occur with significant probability in $\Gamma^{[+]}_1(\Delta)$ is less than 10\% for the multi-CDS hedge, whereas it is 18.9\% for the vanilla hedge (see Fig~\ref{figGammaMulti}).  Thus, using the overall spread in the values of $\Delta$ that occur with significant probability as a risk measure, one finds that the multi-CDS hedge produces a hedged position for the short-protection illiquid CDS that is significantly less risky that produced by the plain vanilla hedge.

For the illiquid long-protection CDS the situation is less clear.  There is an 83.2\% chance of default occurring before the illiquid CDS has matured.  If default does occur before this maturity, it is clear that the riskiness (measured by the criterion introduced in the previous paragraph) of the hedged-position created with the multi-CDS hedge is less than the riskiness of the position created with the plain vanilla hedge.  On the other hand, when there is no default before $T_M$, which occurs with probability 16.8\%, one finds values of $\Delta$ which are $\Delta^{[-]}_0$ = 18.0\% for the the plain vanilla hedge, and $\Delta^{[-]}_0 = 21.0\%$ for the mulit-CDS hedge. Adding in the possibility that default occurs later than the maturity $T_M$ weakens somewhat the case that the multi-CDS hedge produces a less risky position than does the plain vanilla CDS, when hedging a long-protection illiquid position.

Further ways of evaluating the relative effectiveness of the plain vanilla and multi-CDS hedges will be developed below.

\section{Good-Deal Bounds for Bid and Ask Prices}\label{goodDealBounds}
This section takes the point of  view of a dealer who is asked by a customer to take over an illiquid CDS position.  The dealer who accepts such a request is exposed to risk because the illiquid position cannot be perfectly hedged, and wishes to establish criteria for considering the acquisiton of the illiquid position to be a good deal. The subsections below describe a procedure for establishing good-deal bounds for the bid and ask prices of the illiquid CDS in terms of the properties, described in the previous section, of the hedging portfolios that enforce the no-arbitrage bounds on the bid and ask prices of an illiquid CDS.  

A quantity called the expected return on the capital at risk (also called for short, in this article, the expected rate of return) is introduced in subsection~\ref{expectedReturn}.  Large values of the expected rate of return indicate good deals.  Also, the expected rate of return has the attractive property that it tends to plus infinity as the capital at risk tends to zero and the relevant no-arbitrage bound is approched. The dealer characterizes good deals by setting a lower bound on the expected rate of return such that all values of the expected rate of return greater than this lower bound would be considered good deals.  This lower bound on the expected rate of return then leads to a lower bound on the acceptable values for the ask price, and an upper bound on the acceptable values of the bid price.

Subsection~\ref{BAPrices} gives numerical examples of the procedures used to establish the bid and ask prices.  The relationship between the expected rate of return and the bid and ask prices is described in subsubsection~\ref{setExpectedReturn}, a criterion for choosing between two different hedges (e.g. plain vanilla and multi-CDS) is described in subsubsection~\ref{choosing}, and the use of an effective Sharpe ratio in place of the expected rate of return as a good-deal criterion is described in subsubsection~\ref{sharpeRatio}. 

Subsection~\ref{stability} describes how to estimate the robustness of the calculated good-deal bounds for the bid and ask prices of an illiquid CDS with respect to both small and large changes in the assumed physical probability measures.

\subsection{The Expected Rate of Return and the Good-Deal Bounds}\label{expectedReturn}

Now consider the case of an investor who wishes to unwind a unit-notional short-protection illiquid CDS contract.  The dealer contacted by the investor in this respect envisages taking over the short illiquid CDS contract\footnote{This transaction is equivalent to the dealer selling protection to the investor under the terms of the illiquid contract.}, and hedging it with a corresponding long LUB portfolio purchased on the current market for $V^{(+)}$.  The dealer's combined position (short illiquid CDS and long LUB portfolio) then has a non-negative payoff stream with an expected value of $\overline{\Delta}^{(+)}$.  The dealer proposes to charge the investor the amount $V^{(+)}$ to pay for the hedge, and might also be expected to give back to the investor the fraction $\lambda^{(+)}$ of the positive expected payoff of her hedged position.  The fraction $\lambda^{(+)}$ should be positive, otherwise the dealer would make a positive profit with no risk\footnote{This article assumes that counterparty risk has been nullified by sufficiently strong collateral arrangements.  If this is not the case, the procedure of the present paper will represent only a first step in the process of determining bid and ask prices, and the results obtained will have to adjusted to take into account the counterparty risk.\label{counterparty}}.  On the other hand, $\lambda^{(+)}$ should not be greater than unity, otherwise the dealer would have a positive net expected loss on the transaction.  The net payment of the investor to the dealer for this transaction (i.e. the ask price) is $u^{Old(+)} = V^{(+)} - \lambda^{(+)} \overline{\Delta}^{(+)}.$  A similar argument gives the amount $u^{Old(-)}$ that would be paid by the dealer to an investor (i.e. the bid price) for the investor's long protection illiquid CDS contracts.  In summary
\begin{equation}
	u^{Old(\sigma)}(\lambda^{(\sigma)}) = V^{(\sigma)} -(\sigma 1) 
		\lambda^{(\sigma)} \overline{\Delta}^{(\sigma)}, \ \ 0 < 
		\lambda^{(\sigma)} < 1.
	\label{VShort}
\end{equation}
Bid and ask prices that are in the range defined by the restrictions on $\lambda^{(\sigma)}$ in equation~(\ref{VShort}) are said to be potentially acceptable.  A further restriction (see below) of the selection criterion will be necessary to identify the range of prices that are acceptable to the dealer, i.e. those that represent a good-deal for the dealer.  The parameters $\lambda^{(\sigma)}$ determine the unwind values of the illiquid CDS contracts in question.  Because it is not easy to gain intuition about the values of $\lambda^{(\sigma)}$ that would constitute a good deal for the dealer or the investor, these parameters will be replaced by the quantities $R_T^{(\sigma)}$ defined below.

Note that in the special case that there is a CDS on the market having the same maturity as that of illiquid CDS being hedged, the scaling behavior of $u^{Old(\sigma)}$ with $W$ follows from that of $V^{(\sigma)}$ and $\overline{\Delta}^{(\sigma)}$, which are given by (\ref{VBidAskScale}) and (\ref{deltaBar}), respectively.

The random variable giving the present value of the dealer's realized net loss on the transactions just described is, for a given value of $\lambda^{(\sigma)}$, 
\begin{equation}
	L^{(\sigma)}(\tau,\rho) = -\Delta^{(\sigma)}(\tau,\rho) +
	 \lambda^{(\sigma)}\overline{\Delta}^{(\sigma)}.
	\label{Loss}
\end{equation}

It is useful to have a number of quantitative measures of the risk incurred by a dealer who holds a hedged position.  This risk will depend on the agreed upon price 
$u^{Old(\sigma)}(\lambda^{(\sigma)})$ (or, equivalently, the agreed upon value for $\lambda^{(\sigma)}$).  One of the simplest risk measures is $L_{Max}^{(\sigma)}$, the maximum possible loss to the holder of the hedged position, which is given by
\begin{equation}
	 L_{Max}^{(\sigma)} =
	  	\lambda^{(\sigma)}\overline{\Delta}^{(\sigma)}. 
\end{equation}
Since the maximum possible loss is the amount of capital that the dealer must have available to cover losses in a worst case scenario, it is called the capital at risk for the purposes of this article.  Note that, when $\lambda^{(\sigma)}$ is restricted to the range of potentially acceptable values indicated in equation~(\ref{VShort}), the capital at risk is positive.  Because it has a straight-forward economic interpretation and because its use leads to simple analytic formula for quantities used in the process of establishing good-deal bid and ask prices, it has been chosen to be a key variable in future developments. The expected value of the loss, conditional on the loss being greater than zero, $E^{(\sigma)}(L|L>0)$, is another straight-forward single-parameter measure of risk.

Note that the loss probability density $\Gamma_L(L)$ can be found from
\begin{equation}
	\Gamma_L(L) = \Gamma(\lambda \overline\Delta-L)
	\label{GammaL}
\end{equation}
where $\Gamma(\Delta)$ is probability density for $\Delta$ defined at the end of section~\ref{physicalMeasure}. $\Gamma_L(L)$ contains the information necessary to calculate $L_{Max}$, $E^{(\sigma)}(L|L>0)$, as well as other single parameter risk measures such as the mean and variance of $L$ (or of $\Delta$), and thus in principle contains within it a comprehensive view of the risk of a given hedged position. 

Note that pricing at the no-arbitrage bound is characterized by $\lambda^{(\sigma)} = 0$, and hence by the capital at risk $L_{Max}^{(\sigma)} = 0$, in agreement with the idea that there is no risk for a price at the no-arbitrage bound, since the payoffs to the holder of the hedged position are non-negative.  This property of the capital at risk, that it tends to zero when the price for the illiquid CDS tends to that of the no-arbitrage bound, is a desirable property for a risk measure to have, and is found here for the case when hedging is carried out in terms of the porfolio that enforces the no-arbitrage bound.  By way of contrast, the overall spread in the realized present values  of the hedged position, as characterized, for example, by the variance of the random variable $\Delta$, does not have this property.

Another relevant quantity is the dealer's expected profit, called $\overline{P_{FT}}^{(\sigma)}$.  Since the realized profit is the negative of the realized loss, equation~(\ref{Loss}) gives 
\begin{equation}
	\overline{P_{FT}}^{(\sigma)} = (1-\lambda^{(\sigma)})
		\overline{\Delta}^{(\sigma)}.
\end{equation}
Note that the dealer's expected profit is positive when $\lambda^{(\sigma)}$ is restricted to the range of potentially acceptable values indicated in equation~(\ref{VShort}). 

Given that the capital at risk, $L_{Max}^{(\sigma)}$, is positive, the dealer should have a reserve of capital from which this maximum possible loss could be covered, if necessary.  The dealer is entitled to compensation for holding this capital and therefore might wish to calculate a expected rate of return $R_T^{(\sigma)}$ on this capital from the relation
\begin{equation}
	R_T^{(\sigma)} 
		=  \frac{\overline{P_{FT}}^{(\sigma)}}{L_{Max}^{(\sigma)}}
		= \frac{1-\lambda^{(\sigma)}}{\lambda^{(\sigma)}};
		\ \ \text{which implies} \ \ \lambda^{(\sigma)} = 
		\frac{1}{1+R_T^{(\sigma)}}.		
		\label{expectedReturnDef}
\end{equation}
Note that the range of potentially acceptable values of $R_T^{(\sigma)}$ is given by 
\begin{equation}
	0 < R_T^{(\sigma)} < + \infty
\end{equation}
The quantity $R_T^{(\sigma)}$, which is called the expected return on the capital at risk, is also called, for short, the expected rate of return.  

Equations~(\ref{VShort}) can now be rewritten in such a way that the bid and ask prices are determined in terms of the expected rate of return, $R_T^{(\sigma)}$, rather than the parameter $\lambda^{(\sigma)}$.
First define the bounds on the potentially acceptable bid and ask prices by
\begin{eqnarray}
	u^{Old(-)}_{min} & = & V^{(-)}  ;\ \ u^{Old(-)}_{max} = V^{(-)}+ \overline{\Delta}^{(-)};
		\nonumber \\		  
	u^{Old(+)}_{min} & = & V^{(+)}- \overline{\Delta}^{(+)};\ \ u^{Old(+)}_{max} = V^{(+)}.
	\label{bidAskMaxMin}
\end{eqnarray}	
With these definitions, equations~\ref{VShort} can be rewritten as
\begin{eqnarray}
	u^{Old(-)}(R_T^{(-)}) &=& u^{Old(-)}_{min} +(u^{Old(-)}_{max} - u^{Old(-)}_{min})\frac{1}{1+R_T^{(-)}};
				\nonumber \\ 
u^{Old(+)}(R_T^{(+)}) &=& u^{Old(+)}_{min} +(u^{Old(+)}_{max} - u^{Old(+)}_{min})\frac{R_T^{(+)}}{1+R_T^{(+)}}.
	\label{bidAskVsRT}
\end{eqnarray}		
The solution of these equations for the expected rate of return as a function of the bid or ask price is
\begin{eqnarray}
	R_T^{(-)}(u^{Old(-)}) = \frac{u^{Old(-)}_{max} - u^{Old(-)}}{u^{Old(-)} - u^{Old(-)}_{min}};
		\nonumber \\
	R_T^{(+)}(u^{Old(+)}) = \frac{u^{Old(+)} - u^{Old(+)}_{min}}{u^{Old(+)}_{max} - u^{Old(+)}}.
	\label{RTVsBidAsk}
\end{eqnarray}

Note, from equations~(\ref{RTVsBidAsk}), that when the ask price approaches its maximum potentially acceptable value, $u^{Old(+)}_{max}$ (which is the upper no-arbitrage bound) from below, the expected rate of return tends to $+\infty$.  This is because the capital at risk tends to zero at the upper no-arbitrage bound.  Thus, the expected rate of return would be considered to be excessive for an ask price close to its upper no-arbitrage bound.  On the other hand, when the ask price approaches its minimum potentially acceptable value $u^{Old(+)}_{min}$ from above, the expected rate of return tends to zero. For zero expected rate of return, the dealer will have a positive capital at risk equal to $L^{(+)} = \overline{\Delta}^{(+)}$.  However, the dealer would expect to earn a positive return on any capital at risk.  To accomplish this the dealer can decide to accept only those deals that would give an expected rate of return higher than a subjectively chosen positive good-deal lower bound (for the expected rate of return).  This good-deal lower bound on the expected rate of return gives, from equations~(\ref{bidAskVsRT}), a good deal lower-bound on the ask price. The dealer would require the investor to pay more than the good-deal ask-price lower bound in order for the dealer to take over the investor's short-protection position. In so far as the dealer is concerned, there is of course no upper bound on the ask price.  If an investor is happy to pay more than the no-arbitrage upper bound of $u^{Old(+)}_{max}$, the dealer will be happy to accept the arbitrage profit.  The investor, however, may wish to establish a good-deal upper bound on the ask price.  Similarly, the dealer will establish a good deal upper-bound on the bid price: the dealer will buy a long-protection contract from an investor only if the bid price is less that this upper bound. The establishment by the dealer of lower and upper good-deal bounds for the ask and bid prices, respectively, is further illustrated in subsection~\ref{BAPrices} and Fig.~\ref{figExpectedReturn}.

Another possible choice for a quantity for which high values would signal a good deal is the Sharpe ratio.  This ratio has been consider for good-deal indentification in \citet{car01,cer02,coc00}. For the purposes of this article, an analogue of the Sharpe ratio, called the effective Sharpe ratio, is defined as
\begin{equation}
	S_R^{(\sigma)} 
		=  \frac{R_T^{(\sigma)}}{L_{Max}^{(\sigma)}}.
		\label{sharpeRatioDef}
\end{equation}
The use of the effective Sharpe ratio to establish bid and ask good-deal bounds is illustrated in Fig.~\ref{figSharpeRatio}. 	

\subsection{Establishing the Good-Deal Bounds}\label{BAPrices}
\subsubsection{Dealer Sets Expected Return}\label{setExpectedReturn}

\begin{figure} [t!] 
\includegraphics[scale = 0.4]{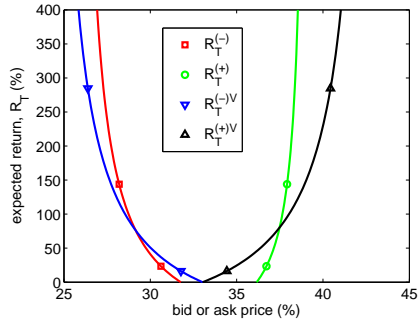}
\caption{Plot of the expected rate of return (shortened to expected return on the vertical axis label), $R_T^{(-)}$ (or $R_T^{(+)}$), versus the bid (or ask) price. A superscript $V$ in the legend indicates results for the vanilla hedge; otherwise, the results are for the multi-CDS hedge.  For the multi-CDS hedge, for bid prices greater than the maximum potentially acceptable bid price of 31.77 \%, and for ask prices less than the minimum potentially acceptable ask price of 36.06 \%, the expected profit and the expected rate of return on capital at risk are both negative.  For the vanilla hedge, the maximum potentially acceptable bid price and the minimum potentially acceptable ask price are both equal to 33.03\%.} 
\label{figExpectedReturn}
\end{figure}

 Subsection~\ref{expectedReturn} described how the dealer could arrive at a good-deal lower bound on the ask price and a good-deal upper bound on the bid price in terms of a good deal lower bound on the expected rate of return.  Fig.~\ref{figExpectedReturn} shows graphically the relationship between the expected rate of return, and the bid and ask prices; and thus also the relationship between the good-deal bounds on the expected rate of return, and the good-deal bounds on the bid and ask prices.

Note from the figure, that for the multi-CDS hedge, a lower bound on the potentially acceptable range for the bid-ask spread is 4.29 \%. It is also clear for the case of the multi-CDS hedge that, if the bid and ask prices were equal, as in equation~(\ref{stdCDSValue}), the dealer would have a negative expected rate of return for either the bid price, or the ask price, or both.
\begin{figure} [t!] 
\includegraphics[scale = 0.4]{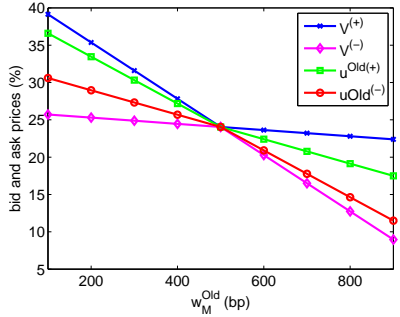}
\caption{The quantities $V^{(+)}$ and $u^{Old(+)}$ are the upper no-arbitrage and good-deal bounds, respectively, on the ask price; while $V^{(-)}$ and $u^{Old(-)}$ are the lower no-arbitrage and good-deal bounds, respectively, on the bid price.  These quantities are plotted versus $w^{Old}$. The multi-CDS hedge was used to obtain these results.} 
\label{figBidAskVsWOld}
\end{figure}

Note also in Fig.~\ref{figExpectedReturn} that, for the vanilla hedge, the bid-ask spread is zero when the required expected rate of return is zero.  There is, however, a significant capital at risk when the expected rate of return is zero, so the dealer who concludes a deal at zero expected rate of return will be in the undesirable position of receiving a zero expected rate of return in compensation for the risk taken on.  Thus, a non-zero bid-ask spread is expected in the case of the vanilla hedge also. It is of interest to contrast this result with the result of the complete-market risk-neutral approach to valuation, which also includes only a single CDS in the hedge, but which gives equal bid and ask prices (as, for example, in equation~(\ref{stdCDSValue})). 

Finally, it is of interest to compare, for the multi-CDS hedge, the no-arbitrage bounds on the bid and ask prices already displayed in Fig.~\ref{VVsWOld}, with the good-deal bounds determined here.  This is done, versus $w^{Old}$, in Fig.~\ref{figBidAskVsWOld}.  The value of $R_T$ used to obtain this plot is given in Table~\ref{SPL}.

For simplicity, the examples used in this article focus mainly on the use of the expected rate of return as a criterion for exstablishing a good deal. However, it is clear (from Fig.~\ref{figExpectedReturn}, for example) that the bid-ask spread could also be used as a good deal criterion.  One could also set minimum values for both the expected rate of return and the bid-ask as the criterion for a good deal.  

\subsubsection{Choosing Between Two Hedges}\label{choosing}

\begin{figure} [t!] 
\includegraphics[scale = 0.4]{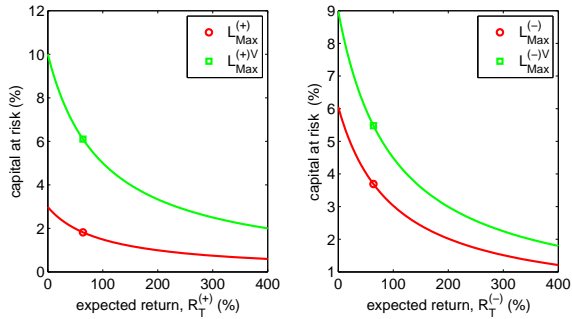}
\caption{Plot of the capital at risk versus the expected rate of return for both the vanilla hedge (superscript V in the legends) and the multi-CDS hedge.  The left (right) hand panel is for the case where the dealer acquires an investor's short (long) illiquid position.} 
\label{figCapitalVsReturn}
\end{figure}

This subsection assumes initially that, when the dealer who takes over an illiquid CDS has a choice of two or more hedges, the dealer will choose the hedge that minimizes the amount of capital at risk, assuming that a definite value for the expected rate of return has been fixed, independently of the consequences that this might have for the bid and ask prices.  Making this choice on the basis of the effective Sharpe ratio is discussed at the end of this subsection.

Fig.~\ref{figCapitalVsReturn} shows the capital at risk plotted against the expected rate of return for both the multi-CDS and the vanilla hedges for input parameters taken from the Standard Parameter List printed at the beginning of  Section~\ref{NABEx}.  Clearly, for any fixed value of the expected rate of return, the capital at risk is significantly reduced by choosing the multi-CDS hedge.  

\subsubsection{The Effective Sharpe Ratio as a Good-Deal Criterion}\label{sharpeRatio} 

Above it was shown how the establishment of a good-deal lower bound for the expected rate of return gave good-deal bounds for the bid and ask prices. The expected rate of return is a quantity for which large values mean good deals.  Similarly, the effective Sharpe ratio, defined by equation~(\ref{sharpeRatioDef}), is a quantity for which large values indicate good deals.  The choice of the effective Sharpe ratio over the expected rate of return adds weight to deals which have a smaller capital at risk for a given expected rate of return. 

To find the dependence of the capital at risk on $\overline{\Delta}^{(\sigma)}$ for a given  effective Sharpe ratio, begin by eliminating $R_T^{(\sigma)}$ from equations~(\ref{sharpeRatioDef}) and (\ref{LMaxOfDeltaSR}), which gives $S_R (L_{Max}^{(\sigma)})^2 + L_{Max}^{(\sigma)} - \overline{\Delta}^{(\sigma)} = 0$.  The positive root of this equation is
\begin{equation}
	L_{Max}^{(\sigma)} = \frac{\sqrt{1+4S_R \overline{\Delta}^{(\sigma)}}-1} 
	{2S_R},
	\label{LMaxOfDeltaSR}
\end{equation}
which shows that, at fixed $S_R$, $L_{Max}$	is a monotonically increasing function of $\overline{\Delta}^{(\sigma)}$.  Thus, for a given effective Sharpe ratio, the preferred hedge is the one with the lowest value of $\overline{\Delta}^{(\sigma)}$, and, as in the previous section, the multi-CDS hedge enforcing the no-arbitrage bounds is preferred relative to the vanilla hedge.

\begin{figure} [t!] 
\includegraphics[scale = 0.4]{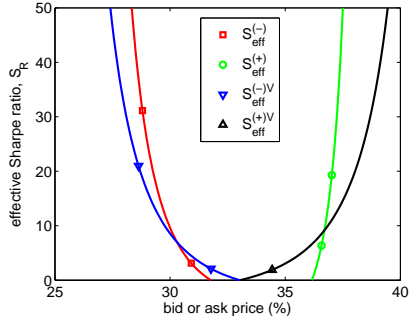}
\caption{This figure is similar to Fig.~\ref{figExpectedReturn} except that the expected rate of return is replaced by the effective Sharpe ration, $S_R$, as the quantity that is selected by the dealer to define the good-deal lower bound.} 
\label{figSharpeRatio}
\end{figure}

Fig.~\ref{figSharpeRatio} shows the effective Sharpe ratio plotted as a function of the bid and ask prices for the same set of input parameters as was used to obtain Fig.~\ref{figExpectedReturn}.  The effective Sharpe ratio can be used in the same way as the expected rate of return to determine good-deal bounds for the bid and ask prices.

\subsection{Robustness of Good-Deal Bid and Ask Prices}\label{stability}
The physical probability measure used to obtain the recommended values $u^{Old(\sigma)}$ of a illiquid CDS must be fixed by the individual or firm carrying out the valuation procedure, based on research to determine realistic physical probability distributions of default times and recovery rates given default.  Thus, the assumed physical probability distributions, while based on research, also depend to a certain extent on judgement.  Confidence in the procedure will be increased, therefore, if it can be shown that  changes in the parameters determining the physical probability distributions, and thus, the distributions themselves, produce only relatively small changes in the recommended values of illiquid CDSs being valued (e.g. see \cite{sta08}) .  This section gives an example Figure showing how the robustness of the bid and ask prices with respect to large variations of the assumed physical probability distributions can be assessed, and also shows that in the example considered, the bid and ask prices are reasonably robust. The results described here are for the multi-CDS hedge that was described in most of the numerical examples above.

\begin{figure} [t!] 
\includegraphics[scale = 0.4]{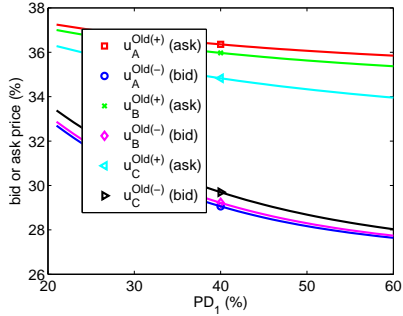}
\caption{Plot, versus $PD_1$, of the bid and ask prices of the illiquid CDS studied in section \ref{goodDealBounds}. The bid and ask prices labelled by the subscript A (or B, or C) have been calculated assuming the recovery rate probability density $\gamma_A(\rho)$ (or $\gamma_B(\rho)$, or $\gamma_C(\rho)$) of Fig.~\ref{recoveryRateDensityABC}. Here $R_T = 25\%$.} 
\label{figuVsPD}
\end{figure}

Fig.~\ref{figuVsPD} describes the dependence of the bid and ask prices on the assumed physical default-time probability distribution and recovery-rate distribution. Note that substantial changes in the assumed physical probability distributions are considered, corresponding to a probability of default within 1 year varying between 20\% to 60\%, and a recovery rate probability density changing from $\gamma_A$ to $\gamma_B$, or to $\gamma_C$, of Fig.~\ref{recoveryRateDensityABC}. (In fact the results shown in Fig.~\ref{figuVsPD} depend only on $\overline{\rho}$ and not on the details of the recovery-rate probability density.)  The other parameter values assumed in obtaining these results are those from the Standard Parameter List given at the beginning of Section~\ref{NABEx}. In going from $\gamma_A$ to $\gamma_B$, the expected recovery rate increases from $\overline{\rho}_A = 20 \%$ to $\overline{\rho}_B = 25 \%$, an increase of 25 \%, while the ask price changes by only about 4 \% of bid-ask spread.  Also, in going from $\gamma_A$ to $\gamma_C$, the expected recovery rate increases by 100 \% (from 20 \% to 40 \%), while the change in the ask price is about 20 \% of the bid-ask spread.  The changes in the bid prices are roughly a factor of 2 smaller than the changes in the ask prices.  Thus, the bid and ask prices are reasonably robust with respect to the choice of the recovery rate probability density and the default time distribution.

In addition, one can easily calculate the numerical sensitivities of the good-deal  bid and ask prices with respect to small changes in the parameters defining the physical probability distributions, such as $PD_1$ for the default time distribution, and the mean and standard deviation parameters that define the recovery rate density. This is an efficient way of examining the robustness of the results with respect to small changes in the physical probability measure.  However, looking at the sensitivities only is not a good way of examining the effect of a large change in the physical probability distributions. In summary, there are procedures for evaluating the robustness of the bid and ask prices for both small and large changes in the physical measure.

\section{Summary and Conclusions}\label{conclusions}
CDS contracts purchased on the CDS market some time ago may have spreads and/or termination dates that do not correspond to CDS contracts on the current CDS market, and are thus illiquid.  Such contracts are valued in this article by considering a transaction in which an investor sells the illiquid contract to a dealer.  Although the dealer hedges the contract as well as possible in terms of CDSs on the current liquid CDS market, she is still left with a hedged position that is risky.  This means that the realized value of the dealer's hedged position is uncertain, and in particular, that the dealer has the possibility to realize a positive loss on the transaction, the maximum value of which is called the capital at risk. 

The article sets up a detailed procedure for arriving at what are called good-deal bounds (from the point of view of the dealer) for the bid and ask prices of the illiquid CDS.  For ask prices greater than a lower ask-price good-deal bound, and for bid prices less than an upper good-deal bid-price bound, the dealer is guaranteed to make an expected rate of retun which is greater than the lower good-deal bound on the expected rate of retun. This lower good-deal bound on the expected rate of retun is set by the dealer as the lowest value of expected rate of return that would be acceptable to her.  Similarly, the good-deal bounds could be described in terms of an analogue of the Sharpe ratio, rather than the expected rate of return.

The implementation of this program is carried out by first setting up a model that can be solved to determine no-arbitrage bounds for the bid and ask prices of the illiquid CDS.  Athough these bounds are too wide to be of use as bid and ask prices for an illiquid CDS, detailed investigation shows that the hedging portfolios that enforce these bounds are useful hedges for the illiquid CDS.  Also, a physical probability measure is established which requires the dealer to specify a probability density for the default times of the reference name (which can be simply done in terms of an  estimate of the probability of default of reference name of the illiquid CDS within one year) as well as a probability density for the recovery rate. Once the hedging portfolios and the physical probability measure have been determined, and the dealer has established a lower good-deal bound for the expected rate of return, the good deal bounds on the bid and ask prices can be calculated. The procedures described have been implemented numerically, and numerical plots illustrating the behavior of a number of important quantities have been included in the article.  Also, a procedure for examining the robustness of the bid and ask prices with respect to a mis-specification of the physical probability measure is described, and its application to a numerical example shows that the good-deal bid and ask prices for that example can be reasonably robust.

The approach allows considerable flexibility in the charaterization of good deals, as the expected rate of return, or the Sharpe ratio, or the maximum possible loss, or the  bid-ask spread, or other measures, either alone or in combination, could be used to characterize good deals.

In the course of this work, a result of general interest in linear programming was obtained.  The result is that in cases where the cost vector in the objective function is random, the solution obtained is unique.


\begin{appendix}
\section{Appendix: Uniqueness and Its Likelihood}\label{uniqueness}

This appendix shows that, if there is a solution to the linear programming problem for which the objective function is $c^Tv$ with the cost vector $c$ random, then the solution is unique.

For simplicity, consider first the two-dimensional case where the objective function is $c^Tv$ with the cost vector $c^T = [u\ 1]$ and $v=[\alpha\ \beta]$, and the optimization problem is the primal problem of (\ref{primalDualLUB}) or (\ref{primalDualGLB}).  The feasible region is a convex two-dimensional polyhedron.  Elementary geometrical arguments (e.g. see \cite{chi10}) show that the solutions to this problem can be of two types. The type of solution is determined by the orientation of the cost vector.  If the cost vector is normal to an edge of the feasible polyhedron and has the right orientation, then every point on that edge gives a solution with the same optimal value, and the solution is not unique.  On the other hand, if the cost vector is not normal to an edge of the feasible polyhedron and there is a solution, then that solution is unique, and the solution point is a corner point of the feasible polyhedron.  In this case, there is a non-zero range orientations of the cost vector (corresponding to a non-zero range of values of $u$) which all have the same corner-point solution. If the feasible region is unbounded (as it is for the problems considered in this paper) then there is the possibility that there is no solution.  If the feasible region is empty (which does not occur for the problems of this paper), there is also no solution.

The upfront price $u$ is a market price.  As such, it can be considered to be a random variable that will depend on supply and demand.  For example, one can consider a practitioner who values the illliquid CDS daily at noon, using the current upfront prices for the market CDSs, which vary randomly from day to day.  Since there is no reason for the market to favor any discrete value for $u$, the probability for finding the random variable corresponding to $u$ in the interval $(u,u+du)$ will have the form ${\cal P}(u)du$, where ${\cal P}(u)$ is a smoothly varying probability density. The probability density $P(u)$ can be conditioned on the value observed for the $u$ on the previous day, or on other relevant parameters. On this basis, the probability  that $u$ has any particular value on a given day (e.g. such that the cost vector $c$ is exactly normal to an edge of the feasible polyhedron) is zero.  Therefore, solutions of the two-dimensional problem that are not unique have zero probability of occurring, whereas those that are unique can occur with a non-zero probability.  Also, there can be a non-zero probability of having no solution.

The result that solutions of the linear programming optimization problem that are not unique have zero probability of occurring can be extended to the general $(K+1)$-dimensional case where $c^T = [u_1,u_2,\dots,u_K,1]$. This follows from a theorem of \cite{man79} which states that,``A linear programming solution is unique if and only if it remains a solution to each linear program obtained by an arbitrary but sufficiently small perturbation of its cost row.''   An arbitrary but sufficiently small variation of the orientation of the cost vector is obtained by making an arbitrary but sufficiently small variation of all of $u_1,\ u_2\ \dots,\ u_K$.  According to the stated theorem, if a solution is unique, the same solution is obtained for arbitrary but sufficiently small variations of all $u_p$'s, and such a solution thus occurs with non-zero probability in the case where all $u_p$'s are considered to be random variables. If the same solution is not reproduced by arbitrary but sufficiently small variations of all $u_p$'s, i.e. if there is a relation such that at least one of the $u_p$'s is determined in terms of the others, then the solution is not unique. Furthermore, a situation in which at least one of the $u_p$'s is determined in terms of the others will have zero probability of occurring in a market  in which each $u_p$ has its own random market value.
\end{appendix}

\end{document}